\documentclass[aps,amsfonts,pre,twocolumn,superscriptaddress,showpacs,eqsecnum]{revtex4-1}
\usepackage{epsfig,amsmath,amssymb,bm,epsf,graphics,psfrag,verbatim,subfigure}
\def\s{\mu}
\def\b{\kappa}
\def\sm{\mu_m}
\def\bm{\kappa_m}

\def\lm{\lambda_m}
\def\ss{\mu_s}
\def\bs{\kappa_s}
\def\vr{\mathbf{r}}
\def\vR{\mathbf{R}}
\def\vu{\mathbf{u}}
\def\uvr{\hat{\mathbf{r}}}
\def\uvepara{\mathbf{e}}
\def\uveperp{\mathbf{e}^{\perp}}

\def\epara{\mathbf{e}}

\def\Prob{p}
\def\Pb{p_{\textrm{b}}}

\def\uv{\mathbf{e}}

\def\DM{\mathbf{D}}
\def\vu{\mathbf{u}}
\def\vq{\mathbf{q}}

\def\vB{\mathbf{B}}

\def\PPV{\mathbf{V}}
\def\GF{\mathbf{G}}
\def\TM{\mathbf{T}}

\def\btwo{\b_c}

\def\bmr{b_m}
\def\lmr{l_m}
\def\br{b}

\def\bmr{b_m}
\def\lmr{l_m}
\def\br{b}

\def\U{\mathbf{U}}

\newcommand{\vl}{\mathbf{l}}

\newcommand{\ev}{\mathbf{e}}
\newcommand{\sv}{\mathbf{s}}

\newcommand{\fv}{\mathbf{f}}
\newcommand{\DMm}{\DM^m}
\newcommand{\GM}{\mathbf{G}}
\newcommand{\GMm}{\GM^m}
\newcommand{\IM}{\mathbf{I}}
\def\a{a}
\def\Pbtwo{p_{b,2}}

\begin{document}

\title{Elasticity of Filamentous Kagome Lattice}
\author{Xiaoming Mao}
\affiliation{Department of Physics and Astronomy, University of
Pennsylvania, Philadelphia, PA 19104, USA }
\affiliation{Department of Physics, University of
Michigan, Ann Arbor, MI 48109-1040, USA }
\author{Olaf Stenull}
\affiliation{Department of Physics and Astronomy, University of
Pennsylvania, Philadelphia, PA 19104, USA }
\author{T. C. Lubensky}
\affiliation{Department of Physics and Astronomy, University of
Pennsylvania, Philadelphia, PA 19104, USA }

\date{\today}

\begin{abstract}
The diluted kagome lattice, in which bonds are randomly
removed with probability $1-p$, consists of straight lines that
intersect at points with a maximum coordination number of four.
If lines are treated as semi-flexible polymers and crossing
points are treated as crosslinks, this lattice provides a
simple model for two-dimensional filamentous networks.
Lattice-based effective medium theories and numerical
simulations for filaments modeled as elastic rods, with
stretching modulus $\mu$ and bending modulus $\kappa$, are used
to study the elasticity of this lattice as functions of $p$ and
$\kappa$. At $p=1$, elastic response is purely affine, and the
macroscopic elastic modulus $G$ is independent of $\kappa$.
When $\kappa = 0$, the lattice undergoes a first-order rigidity
percolation transition at $p=1$. When $\kappa > 0$, $G$
decreases continuously as $p$ decreases below one, reaching
zero at a continuous rigidity percolation transition at $p=p_b
\approx 0.605$ that is the same for all non-zero values of
$\kappa$. The effective medium theories predict scaling forms
for $G$, which exhibit crossover from bending dominated
response at small $\kappa/\mu$ to stretching-dominated response
at large $\kappa/\mu$ near both $p=1$ and $p=p_b$, that match
simulations with no adjustable parameters near $p=1$. The
affine response as $p\rightarrow 1$ is identified with the
approach to a state with sample-crossing straight filaments
treated as elastic rods.

\end{abstract}


\maketitle

\section{introduction}\label{SEC:Intro}

\begin{figure}
	\includegraphics{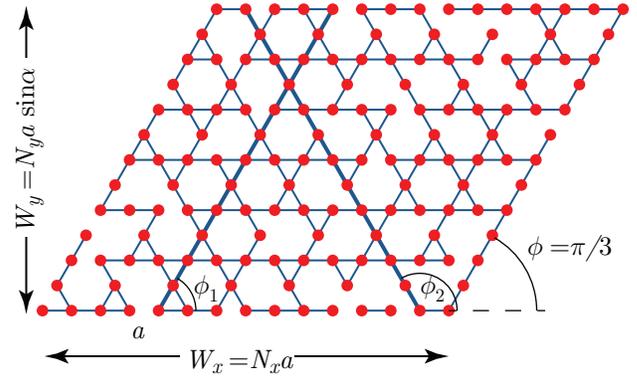}
	\caption{A bond-diluted kagome lattice with $p=0.8$ and $N_x=14$ horizontal bonds and
    $N_y=12$ bonds along the axis at an angle $\alpha = \pi/3$ to the $x$-axis.
    The red sites on the lattice are crosslinks, and
    colinear sequences of bonds are filaments.  Only two filaments, shown with thicker bonds,
    traverse the lattice from top to bottom and support external stresses. One
    these makes an angle $\phi_1 = \pi/3= \alpha$  and one an angle
    $\phi_2 = 2 \pi/3 =2 \alpha$ with the $x$-axis, and the length of
    both is $N_y a$. In more general lattices, sample crossing filaments can
    orient at any angle $\phi$ to the $x$-axis, but the projection of their length onto the
    vertical axis is $W_y$, and thus their length is $L ( \phi )= W_y/(a \sin \phi)$. }
	\label{FIG:diluted_kagome}
\end{figure}

Filamentous networks \cite{Chawla1998} are an important class
of materials in which the interplay between stretching and
bending energies and temperature lead to unique mechanical
properties such as strong nonlinear response and strain
stiffening \cite{GardelWei2004,Storm2005,Onck2005,KangMac2009},
negative normal stress \cite{JanmeyMac2007,KangMac2009},
non-affine response
\cite{Onck2005,Heussinger2006,HeussingerFre2007a,HuismanBar2008},
crossover from non-affine to affine response
\cite{Head2003,Head2003a}, and power-law frequency dependence
of the storage and loss moduli \cite{GardelWei2004a}. They are
a part of such important components of living matter
\cite{Alberts2008,Elson1988,Janmey1990,Kasza2007} as the
cytoskeleton, the intercellular matrix, and clotted blood and
of industrial materials like paper
\cite{LatvaTim2001a,LatvaTim2001b}. Here we introduce the
diluted kagome lattice, shown in Fig.~\ref{FIG:diluted_kagome},
in which elastic rods on nearest-neighbor ($NN$) bonds are
removed with probability $1-p$, as a model for filamentous
networks in two-dimensions. Lines of contiguous and colinear
occupied bonds are identified as filaments, which are modeled,
as in previous work
\cite{Wilhelm2003,Head2003,Head2003a,Heussinger2006,HeussingerFre2007a,HuismanBar2008}
as elastic rods with one-dimensional stretching modulus $\mu$
and bending modulus $\kappa$.

Effective medium theories (EMTs)
\cite{Soven1969,Kirkpatrick1970,Elliott1974} have proven to be
a powerful tool for the study of random systems. We use
recently derived EMTs
\cite{Das2007,Broedersz2010,Mao2011,Das2012} that treat bending
as well as stretching forces in elastic networks to calculate
the shear modulus $G$ as a function of $\mu$, $\kappa$, and
$p$, and derive scaling forms for its behavior near $p=1$ and
near the $\kappa>0$ rigidity threshold at $p=p_b$. We also
calculate $G$ using numerical simulations.  The simulations and
EMTs are in qualitative agreement over the entire range of $p$
and in quantitative agreement near $p=1$. Our results are also
in general agreement with those for the Mikado model
\cite{Wilhelm2003,Head2003,Onck2005}, including, in particular,
a crossover from bending-dominated nonaffine response at low
density (small $p$) to stretching dominated affine response at
high density ($p$ near $1$). Our model, however, highlights
special properties of networks of straight filaments, including
scaling collapse, described analytically by our EMTs, of the
region near $p=1$ where filament length $L$ approaches
infinity, and the underlying origins of affine response in this
limit.

The building blocks of filamentous networks are typically
semi-flexible polymers, characterized by a bending modulus
$\kappa = l_p k_B T$, where $l_p$ is the persistence length
$k_B$ the Boltzmann constant, and $T$ the temperature. Where
filaments intersect, physical or chemical crosslinks bind them
together but do not inhibit their relative rotation. Since only
two filaments cross at a point, each crosslink is connected to
at most four others. The average distance $l_c$ between
crosslinks is less that $l_p$. A remarkably faithful
description of experimentally measured linear
\cite{MacKintosh1995,Piechocka2012} and nonlinear
\cite{Storm2005} elastic response is provided by a model in
which elastic response is assumed to be affine and in which
filamentous sections between crosslinks are treated as
nonlinear central-force springs with a force-extension curve
determined by the worm-like chain model
\cite{MacKintosh1995,MarkoSig1995,Storm2005} augmented by an
enthalpic stretching energy. This simple model, however, misses
some important properties of filamentous networks, including
the existence of a rigidity percolation threshold
\cite{Feng1984,Jacobs1995}, nonaffine response
\cite{Onck2005,Heussinger2006,HeussingerFre2007a,HuismanBar2008},
and crossover from bending-dominated nonaffine response at
small $L/l_c$ \cite{Head2003,Head2003a}, or low frequency
\cite{HuismanBar2010}, to almost affine, stretching dominated
response at large $L/l_c$ or high frequency.

The Mikado model \cite{Wilhelm2003,Head2003a} provides
additional insights into the physics of filamentous networks.
In this model, straight filaments of length $L$ are deposited
with random position and orientation on a two-dimensional flat
surface and crosslinked at their points of intersection.
Filaments are treated as elastic rods with one-dimensional
stretching and bending moduli, $\mu$ and $\kappa$, rather than
as worm-like chains.  Numerical simulations on this model
reveal a rigidity percolation transition from a floppy network
to one with non-vanishing shear and bulk moduli and with
non-affine bending dominated elastic response.  As $L/l_c$ is
increased, response becomes more affine and stretching
dominated, reaching almost perfectly affine response in the
$L/l_c \rightarrow \infty$ limit.  Our EMTs and numerical
simulations yield similar results for the diluted kagome model.

Though the Mikado and the diluted kagome lattice are quite
similar with crosslinks of maximum coordination four and
variable values of the ratio $L/l_c$, they model slightly
different parts of the phase space of possible two-dimensional
filamentous networks. In particular, since the starting point
of the diluted kagome lattice is the full lattice with all
bonds occupied, it necessarily deals directly with the
$L\rightarrow \infty$ limit, which is not generally accessed in
studies of the Mikado model in which $L$ is restricted to be
less than the sample width $W$. The Mikado model is an
``off-lattice" model, whereas the kagome model is lattice
based. The latter property of the kagome model facilitates the
application of lattice-based EMTs
\cite{Soven1969,Kirkpatrick1970,Elliott1974} (modified to
include bending) for all values of $p$.  Finally, In the
Mikado model, the distance between between crosslinks on a
single filament follows a Poisson distribution with no lower
bound, whereas in the kagome lattice, this distance is an
integral multiple of the lattice spacing with a lower bound of
one lattice spacing.  This difference leads, as we shall see,
to a different scaling of the shear modulus in the bending
dominated regime near $p=1$.

As detailed in Apps.~\ref{APP:Asym} and \ref{SEC:AppC}, our
EMTs clearly show that there are three critical points (or
fixed points in the renormalization-group sense): the trivial
empty lattice point at $p=0$ (which we ignore), the
rigidity-percolation point at $p=p_b$, and the $p=1$ full
lattice point.  It also provides analytic scaling relations for
the the shear modulus $G$ as a function of $\mu$, $\kappa$, and
$p$ in the $W \rightarrow \infty$ limit near both $p=p_b$ and
$p=1$. Of particular interest is the behavior near $p=1$, where
the ratio $G/G_0$ of the shear modulus to its $p=1$ value $G_0$
can be expressed as a function of the single scaling variable
\begin{equation}
\tau=\frac{\kappa}{(1-p)^2 \mu a^2} \sim \frac{l_b^2 \langle L \rangle^2}{l_c^4} ,
\end{equation}
where $l_b^2 = \kappa/\mu$ is the bending length $\langle L
\rangle = a/(1-p)$ is the average filament length and $l_c
\approx a$ is the average spacing between crosslinks along a
filament.  As $\tau\rightarrow\infty$, $G$ approaches $G_0$,
but as $\tau\rightarrow 0$, there is  crossover to a bending
dominated, nonaffine regime in which
\begin{equation}
G \propto \frac{\kappa}{l_c^3} \frac{\langle L \rangle^2}{l_c^2} .
\end{equation}
This behavior, which appears in our simulations, has also been
observed in simulations in diluted three-dimensional lattices
of four-fold coordinated filaments
\cite{Stenull2011,BroederszMac2012}. In
Sec.~\ref{SEC:discussion}, we speculate about the reasons for
the same behavior appearing in both two- and three-dimensions.
Huessinger \emph{et al.}
\cite{Heussinger2006,HeussingerFre2007a} developed an
off-lattice EMT for filamentous networks that predicts a
bending-dominated non-affine regime with a different power of
$L/l_c$,  $G \sim(\kappa/l_c^3) (L/l_c)^4$, than the one we
predict.  The origin of this different scaling is the
absence of a lower cutoff in the Poisson distribution of the
distance between crosslinks in the Mikado model compared to the
fixed cutoff equal to the lattice spacing in the kagome model.
The analysis of Refs.~\cite{Heussinger2006,HeussingerFre2007a}
yields the kagome lattice scaling law if the probability
distribution for distances is replaced by one with a fixed
lower cutoff.

\begin{figure}
\includegraphics[width=.42\textwidth]{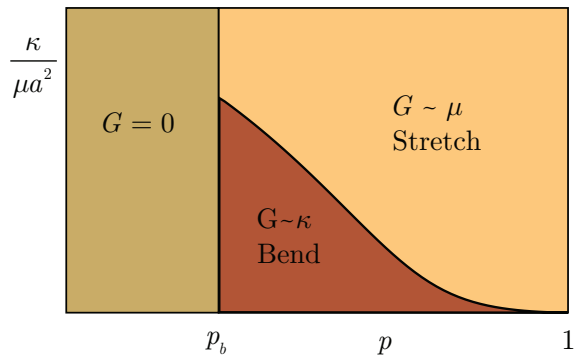}
\caption{Schematic phase diagram for the diluted kagome lattice
showing the bending-dominated region in which $G \sim \kappa$ and the
stretching dominated region in which $G\sim \mu$. This boundary between stretching-
and bending-dominated response was calculated using a formula that interpolates
between the analytic EMT expressions for $G$ near $p=p_b$ and near $p=1$. }
\label{FIG:phasediagram}
\end{figure}

Our numerical simulations  follow the EMT prediction very
closely with {\em no} adjustable parameters near $p=1$ if
$\kappa$ is not too large. For really small values of
$\kappa/\mu a^2$, finite-size effects become important, and
simulations can be fit with a combination of the exactly
calculable finite size results at $\kappa =0$ and the EMT
scaling form valid at infinite $W$. Near $p=p_b$, simulations
are consistent with a scaling function of the form of
Eq.~(\ref{EQ:scaling-2}) but with $t$ approximately $0.2$
rather than $1$ and with different values of $c_1$ and $c_2$.

Under periodic boundary conditions, the kagome lattice has
exactly $z=4 = 2d$ neighbors per site, where $d$ is the spatial
dimension.  If sites interact only via central-force springs on
bonds and not by bending forces along filaments, the lattice is
{\em isostatic} in that it has exactly enough internal forces
to eliminate zero modes according the simple Maxwell count
\cite{Maxwell1864}: $N_0 = dN - N_B = (d-\frac{1}{2} z) N$,
where $N_0$ is the number of zero modes and $N_B = \frac{1}{2}
z N$ is the number of bonds. Under free rather than periodic
boundary conditions, sites on the boundaries have fewer than
four neighbors, and there is an overall bond deficiency of
$\Delta N_B = 2 N - N_B \sim \sqrt{N}$.  As a result, there are
of order $\sqrt{N}$ zero ``floppy" internal modes of zero
energy in the absence of bending forces. Under periodic
boundary conditions, the simple Maxwell count leads to $N_0 =
0$, but the total number of zero modes (as calculated by the
dynamical matrix, whose normal modes are the ``infinitesimal"
modes that result when the elastic energy is truncated to
harmonic order), like the number under free boundary
conditions, is of order $\sqrt{N}$ rather than zero
\cite{Souslov2009}.  This discrepancy is a result of the
existence of lattice configurations that can support stress in
bonds while maintaining a zero force at each note
\cite{Sun2012}. Associated with each such configuration
\cite{Calladine1978}, called a ``{\em state of self stress}",
is an additional infinitesimal zero mode so that the Maxwell
count becomes $N_0 = dN - N_B + S$, where $S$ is the number of
states of self stress.  Thus, in the undiluted central-force
kagome lattice under periodic boundary conditions, $N_0=S$ is
simply the number of states of self-stress.  These states endow
the undiluted central-force kagome lattice with affine response
and non-zero bulk and shear moduli, proportional to the bond
spring constant, and, thus, determine the approach to affine,
stretching-dominated response as $p\rightarrow 1$ in the
diluted lattice with nonzero $\kappa$.  In addition, the
central-force zero modes associated with the states of self
stress also control the form of the effective-medium equations
and the scaling forms they predict. The addition of bending
forces lifts all but the two trivial zero modes of rigid
translation to finite frequency. This fact plays a central roll
in our EMT and is ultimately responsible for the form of the
scaling function for $G$ near $p=1$.

We are ultimately interested in three-dimensional filamentous
networks, which are subisostatic with $N_0=3N - 2N = N$ because
their maximum coordination number, like the that of
two-dimensional networks, is four. If constituent filaments are
straight, these networks have a state of self-stress for each
filament in the undiluted limit, and their elastic moduli are
non-zero, proportional to the bond-spring constant, and
independent of $\kappa$ \cite{Stenull2011}. Associated with
each state of self-stress, there is an additional zero mode,
but as in the kagome lattice, the addition of bending forces
raises all but the three trivial zero modes of uniform
translation to finite frequency and stabilizes the lattice. We
will argue in Sec.~\ref{SEC:discussion} that these properties
are the likely origin of the very similar behavior, seen in
simulations, of the shear modulus in $2$- and $3$-dimensional
networks with maximum coordination of four.

This paper consists of five sections and three appendices.
Section \ref{Sec:filnetworks} compares and contrasts the Mikado
and kagome models and demonstrates how straight,
sample-traversing filaments with the energy of an elastic beam
produce affine elastic response. Section \ref{Sec:filnetworks}
also defines the lattice model that we use. Section
\ref{SED:EMT-OR} describes EMT procedures and presents their
results.  Section \ref{SEC:numericalSimulations} presents the
results of our numerical simulations and compares them with
those of the EMTs. Section \ref{SEC:discussion} reviews our
results and speculates about application of our two-dimensional
calculations to three-dimensional systems. There are three
appendices: Appendix \ref{SEC:APP1} discusses general
properties of the EMT dynamical matrix. Appendices
\ref{APP:Asym} and \ref{SEC:AppC} present details of the
solutions to the EMT equations near $p=1$ and $p=p_b$ for the
bending EMTs described in Refs.~\cite{Broedersz2010,Mao2011}
and in Refs.~\cite{Das2007,Das2012}, respectively.

\section{Filamentous networks and the diluted kagome
lattice\label{Sec:filnetworks}}

Networks composed of straight filaments with elastic-rod
energies have special properties. In this section, we will
explore some general properties of these networks before
setting up the energy for the kagome lattice.

\subsection{Elastic filamentous networks}\label{SEC:elasticfil}

The elastic energy of a one-dimensional elastic beam with
stretching modulus $\mu$ and bending modulus $\kappa$ is
\begin{equation}
E_{\text{fil}} = \frac{1}{2} \mu \int_0^L ds \left( \frac{du(s)}{ds} \right)^2
+ \frac{1}{2} \kappa \int_0^L ds \left( \frac{d\theta(s)}{ds} \right)^2 ,
\end{equation}
where $u(s) $ is the local longitudinal displacement and
$\theta(s)$ the local angle of the tangent to the filament at
the point $s$. The usual practice is to treat the stretching
modulus as independent of $\kappa$ even though at nonzero
temperature the spring constant for the entire filament has an
important entropic component that depends on $\kappa$ .

The spring constant for a filament section of length $l$ is $k
= \mu l^{-1}$. This leads to affine response in linear
elasticity for any lattice in any dimension consisting of
sample-traversing straight filaments with a sufficient number
of orientations to ensure stability with respect to all strain
\cite{Heussinger2006,HeussingerFre2007a,GurtnerDur2009}. To see
this, consider a crosslink (node of the lattice) at the origin
on a filament parallel to the unit vector $\ev$, and let it be
connected to two other crosslinks on the same filament at
respective positions $\sv_1 =  l_1 \ev$ and $\sv_1 = -l_2 \ev$
relative to the orign. Under a uniform, affine deformation, the
relative positions $\sv_a$, $a=1,2$ transform to $ \Lambda\cdot
\sv_a \equiv \sv_a + \eta \cdot \sv_a$ , where $\Lambda = I +
\eta$ is the deformation tensor with components $\Lambda_{ij} =
\delta_{ij} + \eta_{ij}$. The forces that crosslinks $1$ and
$2$ exert on the origin are then
\begin{equation}
\fv_1 = k(l_1) (l_1 e_i \eta_{ij} e_j) \ev ; \,\,\,
\,\,\, \fv_2 =-
k(l_2) (l_2 e_i \eta_{ij} e_j) \ev ,
\end{equation}
where $i$ and $j$ run over $x,y$ and the summation convention
is understood. The sum of these forces is zero because $k(l) =
\mu/l$. The same analysis applies to any site and filament in
the lattice. Under affine distortions, filaments do not bend,
so the energy of the affine distortion depends only on the
central force and does not depend on $\kappa$. Thus, under
affine distortions of sample traversing filaments, the force on
every intermediate crosslink is zero, and non-affine
distortions are not introduced:  the response is affine and
independent $\kappa$. When the lattice is diluted, not every
crosslink is connected to two others, the above cancelation
does not occur, and the result is nonaffine response with a
bending component.

With these observations, we can calculate the shear modulus
of any network of straight filaments with stretching energy
only, i.e., $\kappa = 0$.  For simplicity, we restrict our
attention to two-dimensional systems in a rectangular- or
rhombus-shaped box with base-length $W_x$ and height $W_y$ as
shown in Fig.~\ref{FIG:diluted_kagome}, and we consider only
shear deformations in which the only nonvanishing component of
$\eta$ is $\eta_{xy}$. In this case, only those filaments that
cross from the bottom to the top of the sample will contribute
to the shear modulus.  The length of such a sample traversing
filament oriented along a unit vector $\ev(\phi) = (\cos \phi,
\sin \phi)$ making an angle $\phi$ with the $x$-axis is
$L(\phi) = W_y/(a \sin \phi)$. Under a shear deformation
$\eta$, its length will change by $\delta L= L e_i(\phi)
\eta_{ij} e_j(\phi)$ to lowest order in $\eta$. Since the
spring constant of a filament of length $L$ is $\mu/L$, the
elastic energy of the filament is
\begin{equation}\label{Eq:E-fil}
E_{\text{fil}} (\phi) = \frac{1}{2} \mu W_y
\frac{[e_i(\phi) \eta_{ij} e_j(\phi)]^2 }{\sin \phi}.
\end{equation}
and the total energy from all filaments is $E_{\text{tot}}
=(1/2) N W_y\mu \langle [e_i(\phi) \eta_{ij}
e_j(\phi)]^2/\sin\phi\rangle$, where $N$ is the total number of
sample traversing filaments and $\langle \cdots \rangle$
signifies an average over the orientation angles of the
filaments.

The linearized energy density, $E/W_x W_y$, is
\begin{align}
\label{harmicElasticEnDens}
f = \textstyle{\frac{1}{2}} \, C_{ijkl} \, u_{ij} u_{kl}    \, ,
\end{align}
where $C_{ijkl}$ are the components of the elastic constant or
elastic modulus tensor and $u_{ij} = \textstyle{\frac{1}{2}}
(\eta_{ij} + \eta_{ji})$  are the components of the linearized
strain tensor $\underline{\underline{\hat{u}}}$. In the
isotropic limit, the energy density reduces to
\begin{eqnarray}\label{EQ:CONT}
    f =  \frac{1}{2} B \big(\textrm{Tr}\underline{\underline{u}}\big)^2 +
    G \textrm{Tr}\underline{\underline{\hat{u}}}^2 ,
\end{eqnarray}
where $\underline{\underline{\hat{u}}}$ is the traceless part
of  $\underline{\underline{u}}$, and $G =  C_{xyxy}$ and $B =
\textstyle{\frac{1}{2}} (C_{xxxx} +  C_{xxyy})$ are the shear
and bulk moduli, respectively. These relations along with
Eq.~(\ref{Eq:E-fil}) for the case in which the only
nonvanishing component of $\eta_{ij}$ is $\eta_{xy}$ lead to
\begin{equation}
G = \frac{1}{4}\frac{\mu N}{W_x}\left\langle \frac{\sin^2 2 \phi}{\sin\phi}\right\rangle
\end{equation}
for the shear modulus of a network of elastic rods with
stretching modulus $\mu$.  Note that rods parallel to the
$x$-axis ($\phi=0$) do not contribute to $G$. In the kagome
lattice, bottom-to-top traversing filaments all have length
$L_y = W_y /\sin (\pi/3)$ and their angles with respect to the
$x$-axis are restricted to $\phi = \pi/3, 2 \pi/3$ for which
$\sin \phi = \sin 2 \phi = \sqrt{3}/2$. There are $W_x/a$ sites
along the $x$ axis from which a single filament aligned either
along $\pi/3$ or $2 \pi/3$ can emerge, so $N$ is simply
$q_{\text{cross}} W_x/a$, where $q_{\text{cross}}(p)$ is the
probability a filament emerging from a given site along the
$x$-axis traverses the sample.  For simplicity, we take $L_y=W$
in which $q_{\text{cross}}(p,W)= p^{W/a}$. When $p=1$, all
bonds are occupied, and the shear modulus of the undiluted
kagome lattice is $G_0 = (\sqrt{3}/8)(\mu/a)$. When $p<1$ and
$\kappa=0$, there is a nonzero stretching contribution to $G$
in finite samples:
\begin{equation}\label{EQ:Gstr}
G_{\text{str}}(\mu,p,W)  \equiv  G(\mu, \kappa = 0, p, W)
 =  q_{\text{cross}} (p,W) G_0 .
\end{equation}
In the $W\rightarrow \infty$ limit, the probability of sample
traversing filaments vanishes for any $p<1$, and
$G_{\infty}(\mu,\kappa,p) = \lim_{W \rightarrow \infty}
G(\mu,\kappa,p,W)$ is zero at $\kappa=0$ for all $p<1$.  Thus,
as $\kappa \rightarrow 0$, $G_{\infty}(\mu,\kappa,p)$ must
become smaller than $G_{\text{str}}(\mu,p,W) $, and for
sufficiently small $\kappa$ and $p$ near one,
$G(\mu,\kappa,p,W) \approx G_{\text{str}}(\mu,p,W)$, and, as we
will show in Sec.~\ref{SEC:numericalSimulations}, the simple
interpolation formula
\begin{equation}
G(\mu,\kappa,p,W ) \approx \max \{G_{\infty}(\mu,\kappa,p),G_{\text{str}}(\mu,p,W)\}
\label{EQ:finite-size}
\end{equation}
provides and excellent description of the simulation data with
the EMT form for  $G_{\infty}$ near $p=1$ where the finite size
corrections are the most important.

\subsection{Kagome lattice energy}
\begin{figure}
\includegraphics{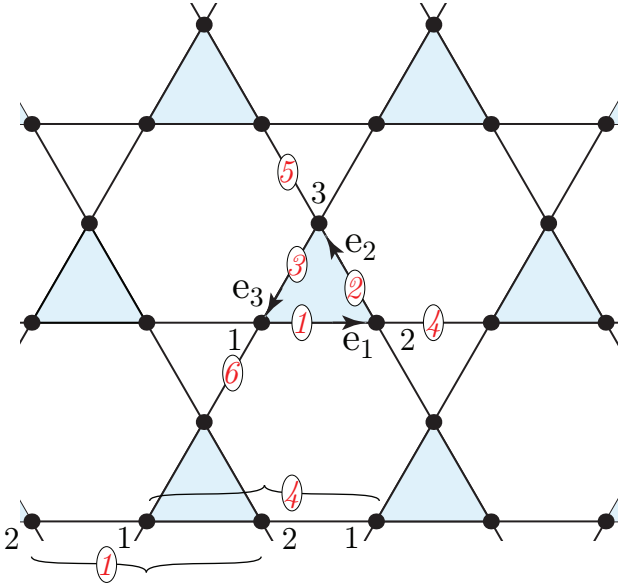}
\caption{A section of a kagome lattice showing lattice directions
$\epara_1$, $\epara_2$, and $\epara_3$. Sites and $NN$ bonds in our
choice for the unit cell labeled $1$, $2$,
and $3$ and  $1-6$ in italics, respectively. Two of the six $NNN$ bonds in a unit cell
are shown at the bottom of the figure.  The other four have a similar
configurations aligned along $\epara_2$ and $\epara_3$ rather than $\epara_1$.}
\label{fig:simpleKagome}
\end{figure}

The kagome lattice has three sites per unit cell, which we take
to be the three sites, labeled $1$, $2$, and $3$, on an
elemental triangle in Fig.~\ref{fig:simpleKagome}. All bonds in
the lattice are parallel to one of the vectors, $\uv_1$,
$\uv_2$, and $\uv_3$, specifying the direction of bonds in the
unit elemental triangle. We label each site by an index $\ell =
(\vl,\sigma)$ where $\vl = (l_1,l_2)$ labels the  position of
site $1$ in the unit cell at dimensionless (i.e., taking $a=
1$) position $\vr_{\vl} = 2(l_1 \uv_1 + l_2 \uv_2)$, where the
factor of two arises because the separation between unit cells
is twice the bond length. The index $\sigma= 1,2,3$ labels the
site within the unit cell, and the position of site $\ell$ is
$\vr_{\ell} = \vr_\vl + \sv_{\sigma}$, where $\sv_1=0$, $\sv_2
= \uv_1$, and $\sv_3=-\uv_3$. When the lattice is distorted,
$\vr_\ell$ maps to a new position $\vR_\ell = \vr_\ell +
\vu_\ell$, where $\vu_\ell$ is the displacement vector of site
$\ell$.  The stretching energy of a bond $\langle \ell,
\ell'\rangle$ connecting nearest-neighbor ($NN$) sites $\ell$
and $\ell'$ is $(1/2)(\mu/\a ) (\vu_{\ell\ell'} \cdot
\uvr_{\ell\ell'})^2$ where $\vu_{\ell\ell'} =
\vu_{\ell'}-\vu_\ell$ and $\uvr_{\ell\ell'}$ is the unit vector
parallel to $\vr_{\ell'}-\vr_\ell$.  Bending energy is
determined by angles $\theta_{\ell\ell,\ell''}$ between  $NN$
contiguous bonds $\langle \ell,\ell'\rangle$ and $\langle
\ell', \ell''\rangle$ parallel to one of the lattice directions
$\uv_n$. To harmonic order, $\theta_{\ell\ell'\ell''} =
(1/\a)(\vu_{\ell\ell'}-\vu_{\ell'\ell''})\cdot \uveperp_n$,
where $\uveperp_n$ is the vector perpendicular to $\uvepara_n$,
which specifies the directions of both bonds $\langle \ell,
\ell'\rangle$ and $\langle \ell',\ell''\rangle$,  according the
right hand rule about the outward normal to the plane. The
bending energy arising from a nonzero
$\theta_{\ell\ell'\ell''}$ is
$(1/2)(\kappa/\a^3)(\theta_{\ell\ell'\ell''})^2$, and the
harmonic energy for our latticed is thus
\begin{subequations}\label{EQ:Hami}
\begin{eqnarray}
		E &=& E_{s}+E_{b} \\
    E_{s}(\mu) &=& \frac{1}{2} \frac{\mu}{\a} \sum_{\langle \ell,\ell'\rangle}
    g_{\ell, \ell'} (\vu_{\ell\ell'}\cdot \uvr_{\ell\ell'})^2 , \label{EQ:ENST}\\
    E_{b}(\kappa) &=& \frac{1}{2} \frac{\kappa}{\a}
    \sum_{\ell,\ell',\ell''} g_{\ell,\ell'} g_{\ell',\ell''}( \theta_{\ell\ell'\ell''})^2\\
    &=& \frac{1}{2} \frac{\kappa}{\a^3} \sum_{\langle \ell,\ell',\ell''\rangle}
    g_{\ell,\ell'} g_{\ell',\ell''}\big\lbrack (\vu_{\ell\ell'}-\vu_{\ell'\ell''})\nonumber
    \cdot \uveperp_n\big\rbrack^2, \label{EQ:ENBE}
\end{eqnarray}
\end{subequations}
where $g_{\ell,\ell'} = 1$ if the bond connecting sites $\ell$
and $\ell'$ is occupied and $g_{\ell,\ell'} = 0$ otherwise and
where the sum in $E_b(\kappa)$ is over all continuous triples
of sites $\ell,\ell',\ell''$ along each filament. The
combination $\mu/\a$ is the stretching elastic constant, and
the combination $\kappa/\a^3$ is the bending constant for
contiguous pairs of bonds.

\section{EMT: Overview and Results}\label{SED:EMT-OR}

Effective medium theory
\cite{Soven1969,Kirkpatrick1970,Elliott1974} is a
well-established approximation for calculating properties such
as electronic band structure and phonon or magnon dispersions
of random media. Various formulations of EMT exist, but all
replace the random medium with a homogeneous one whose
parameters (such as $NN$ hopping strength or spring constant)
are determined by a self-consistent equation.  Here we use the
formulation, presented in greater detail in
App.~\ref{SEC:APP1}, in which self-consistency equations for
the effective-medium parameters are determined by requiring
that the average multiple-scattering potential or $T$-matrix
associated with a single random bond (or group of bonds) in the
homogeneous effective medium vanishes.

The development of an EMT for elastic networks with both
stretching and bending forces presents challenges not
encountered in networks with stretching forces only. In our
system, stretching forces are associated with single $NN$
bonds, but bending forces are associated with contiguous pairs
of $NN$ bonds that couple next-nearest-neighbor ($NNN$) sites
in what we call  \emph{phantom} $NNN$ bonds that only exist if
both its constituent $NN$ bonds are occupied.  Thus, the
replacement of a single $NN$ bond, which we will refer to as
the replacement bond, in an effective medium affects not only
the stretching energy of that bond but also the bending of the
two $NNN$ ``phantom" bonds containing that bond. There are
currently two versions
\cite{Das2007,Broedersz2010,Mao2011,Das2012} of EMT on lattices
with bending energies, which we will refer to as EMT I and
EMT II, respectively, that deal with this problem in different
ways. In both methods, the effective medium is characterized by
homogeneous stretching and bending moduli $\mu_m$ and
$\kappa_m$ respectively.

In EMT I \cite{Broedersz2010,Mao2011}, the replacement bond has
a stretching modulus $\mu_s$ and a bending modulus $\kappa_s$
that take on respective values $\mu$ and $\kappa$ if the bond
is occupied and $0$ if the bond is not occupied. Thus, the
probability distribution for $\mu_s$ and $\kappa_s$,
\begin{equation}\label{EQ:P-I}
P(\mu_s, \kappa_s) = p \delta (\mu_s-\mu) \delta (\kappa_s-\kappa)
+(1-p) \delta(\mu_s) \delta(\kappa_s) ,
\end{equation}
exhibits strong correlation between the values of $\mu_s$ and
$\kappa_s$. As discussed in detail in Ref.~\cite{Mao2011}, the
bending constant of the $NNN$ phantom bonds containing the
replacement bond is calculated assuming it is composed of two
elastic beams connected in series, one with the bending modulus
$\kappa_m$ of the effective medium and one with the bending
modulus $\kappa_s$ of the replacement bond. This leads to a
bending constant, $\kappa_c$, for both $NNN$ bond containing he
replacement bond, that is a nonlinear function of $\kappa_s$
and $\kappa_m$:
\begin{align}
	\btwo(\bs) = 2\left(\frac{1}{\bs}+\frac{1}{\bm}\right)^{-1} .
\end{align}
The perturbation to the effective medium arising from the
replacement bond then consists of a stretching energy on that
bond with spring constant $(\mu_s-\mu_m)/\a$ and bending
constants on the two $NNN$ bonds of $(\kappa_c-\kappa_m)/a^3$.
It turns out, as discussed more fully in
Refs.~(\cite{Broedersz2010,Mao2011}) and in
App.~\ref{APP:Asym}, that the EMT equations in Method I do not
close unless an additional term, with coupling constant
$\lambda_m$, coupling the angles on neighboring $NNN$ bonds
along a single filament is added to the effective medium
energy.  This leads to an additional term in the perturbation
arising from the replacement bond, with coupling constant
$-\lambda_m/a^3$, that couples the two angles on the two $NNN$
bonds containing the replacement bond.  Thus this EMT is
characterized by three parameters $\mu_m,\kappa_m$, and
$\lambda_m$.

In EMT II \cite{Das2007,Das2012}, the phantom $NNN$ bonds
carrying the bending energy are elevated to the status of real
bonds that exist whether or not the $NN$ bonds of which they
are composed are occupied.  This leads to the great
simplification that stretching and bending are effectively
decoupled, and there is no necessity of introducing
$\lambda_m$. There are separate probability distributions for
the stretching modulus of the $NN$ replacement bond and for the
bending modulus of the $NNN$ replacement bond.  However, since
the bending modulus of a $NNN$ bond is zero unless both $NN$
bonds comprising it are occupied, the probability that the
$NNN$ is present with a bending modulus $\kappa$ is chosen to
be $p^2$, the probability that any two given $NN$ bonds are
occupied:
\begin{subequations}
\begin{eqnarray}\label{EQ:P-II}
P(\mu_s) & = & p \delta(\mu_s-\mu) + (1-p) \delta (\mu_s) \\
P(\kappa_s) & = &  p^2 \delta(\kappa_s-\kappa) + (1-p^2) \delta (\kappa_s) .
\end{eqnarray}
\end{subequations}

\begin{figure}
	\includegraphics[width=.45\textwidth]{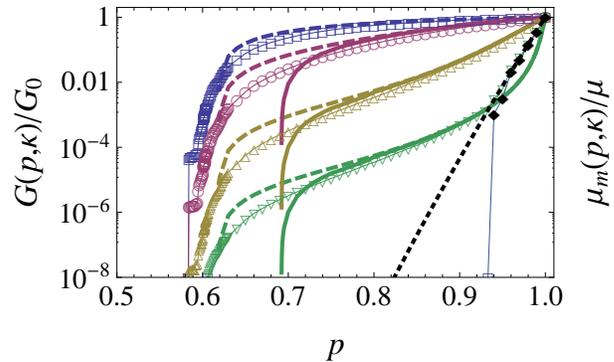}
	\caption{Comparison between the shear modulus (normalized by the shear
    modulus $G_0$ at $p=1$, so it is equivalent to $\mu_m/\mu$) obtained from our numerical
    simulations (data points):   EMT I (solid lines) and EMT II (dashed lines).
    Different colors mark different values of $\b/(\s a^2)$, and from top to bottom
    the corresponding values of $\b/(\s a^2)$ are $1$, $10^{-2}, 10^{-4}, 10^{-6}$, $0$.
    The dotted line indicates the finite size correction predicted
    by Eq.~(\ref{EQ:finite-size}) As $p$ increases toward $1$
    the simulation curve for $\b/(\s a^2)=10^{-6}$
    follows first the EMT curves and then the finite-size $\kappa = 0$ curve beyond the
    point at which the latter curves meet in accord with Eq.~(\ref{EQ:finite-size}).  For the values of $\b/(\s a^2)>10^{-6}$, the EMT solution is larger that the $\kappa = 0$ finite-size result, and thus the simulation data follow the EMT curves.}
	\label{FIG:Compare}
\end{figure}

The predictions of these two methods are qualitatively similar:
they both yield a rigidity percolation threshold, below which
the lattice loses rigidity, and a smoothly varying shear
modulus approaching the pure lattice value of $G_0$ as
$p\rightarrow 1$. Figure \ref{FIG:Compare} plots values of $G(k,\kappa)$ calculated using both EMT methods for different values of $\kappa/(\mu a^2)$ and $p$ along with the results of simulations.   
The $\kappa = 0 $ simulation curves follows the finite-size result of Eq.~\eqref{EQ:Gstr}. 
The $\kappa/(\mu a^2)= 10^{-6}$ curve follows the  EMT curve, which corresponds to infinite $W$, with increasing $p$ and then the $\kappa = 0$ finite size curve after the two cross in accord with Eq.~\eqref{EQ:finite-size}.
As is the case for the triangular
lattice \cite{Das2012}, EMT II predicts a value of $p_b$ that
is close to that measured in simulations, whereas EMT I
predicts a considerably larger value.  In addition, for values of $\kappa/(\mu a^2)$ beyond $0.1$, the EMT I numerical solution ceases to exist for small $p$, and as a result the $\kappa/(\mu a^2)=1$ curve is not included for EMT I in Fig.~\ref{FIG:Compare}.
For $p\gtrsim0.7$ and $\b/(\s a^2)>10^{-6}$, the
two EMT curves and the simulation curves are essentially indistinguishable, and near $p=1$, they become analytically identical
with the effective medium modulus $\mu_m$ satisfying a scaling
form
\begin{align}\label{EQ:muscale}
	\mu_m = \mu \cdot \Psi(\tau)
\end{align}
where the scaling variable is
\begin{align}\label{EQ:tau}
	\tau = \frac{\kappa/(\mu a^2)}{(1-\Prob)^2}
	=\frac{\kappa}{\mu a^4} \langle L \rangle ^2
\end{align}
and the scaling function is
\begin{align}\label{EQ:ScalFunc}
	\Psi(\tau) &= (-1+\sqrt{1+A \tau})^2 / (A \tau) \, .
	\end{align}
with the constant $A=[40(1-\sqrt{2/3})]^2\simeq 53.9$. This
scaling function can be expanded in the following limits
\begin{align}
	\Psi (\tau) \simeq \left\{ \begin{array}{ll}
		1 & \textrm{if $\tau \gg 1$} \\
		A \tau/4 & \textrm{if $\tau \ll 1$}
		\end{array}  \right .
\end{align}
Therefore, for the case of $\tau\gg 1$ corresponding to $\Prob$
near $1$ so that $\kappa/(\mu a^2) \gg (1-\Prob)^2$, the shear
modulus is approximately
\begin{align}
	G =\frac{\sqrt{3}}{8}\frac{\mu_m}{a}
    \approx \frac{\sqrt{3}}{8}\frac{\mu}{a}= G_0 ,
\end{align}
indicating that the macroscopic elastic response of the network
is dominated by the stretching stiffness of the filaments and
reaches the affine pure lattice limit as $p \rightarrow 1$, and
we shall call this the \lq\lq stretching dominated\rq\rq
elastic regime.

\begin{figure}
\includegraphics[width=8.4cm]{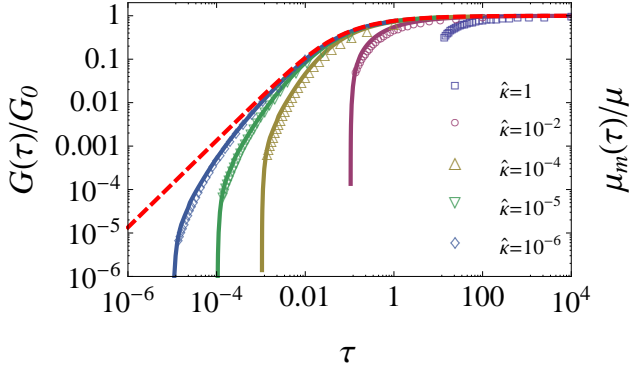}
\caption{(Color online) Scaling behavior near $p=1$ for different values of $\hat{\b}\equiv\b/(\s a^2)$. The data
points for the shear modulus stem from our simulations for $200^2$ unit cells.
The continuous curves depict our numerical solutions for $\mu_m$ of the EMT
equations for the corresponding values of $\kappa$. The dashed red curve
shows the asymptotic scaling function predicted by the EMT, see Eq.~(\ref{EQ:ScalFunc}).
As expected, the simulation data and the numerical solution of the EMT equations
deviate from this scaling function as $1-p$ becomes large.}
\label{fig:scalingNearP=1}
\end{figure}

In the other limit $\tau\ll 1$ corresponding to smaller
$\kappa$ or larger $1-\Prob$ so that $\kappa/(\mu a^2) \ll
(1-\Prob)^2$, we have the shear modulus
\begin{align}
	G \simeq A\frac{\sqrt{3}}{32}\frac{\kappa}{a^3 (1-\Prob)^2}=
A\frac{\sqrt{3}}{32}\frac{\kappa}{a^3}\frac{L^2}{a^2} ,
\end{align}
indicating that the macroscopic elastic response of the network
is dominated by the bending stiffness of the filaments, and we
shall call this the \lq\lq bending dominated\rq\rq  elastic
regime.  The EMT solution for $p$ near one is plotted in terms
of the scaling variable $\tau$ together with the asymptotic
scaling function $\Psi(\tau)$ in Fig.~\ref{fig:scalingNearP=1}.
Because the asymptotic solution~\eqref{EQ:ScalFunc} assumes
small $(1-p)$, it requires very small $\kappa/(\mu a^2)$ to
make the regime $\Psi(\tau)\simeq A \tau/4$ visible. We
conclude from this that the elasticity of the network is
bending dominant as long as
\begin{align}
	\kappa/(\mu a^2) \ll (1-\Prob)^2 .
\end{align}

As discussed in Sec.~\ref{SEC:Intro}, near the rigidity
threshold, $G(p,\kappa)$ vanishes as $\mu (p-p_b)^t$, 
where $t=1$ and $p_b = 0.6920$ for EMT I and $p_b = 0.6180$ for
EMT II, times a scaling function of $\bmr=\kappa/(\mu a^2)$
that is a constant at large $\bmr$ and proportional to $\bmr$
at small $\bmr$. Thus, at sufficiently small $\bmr$ for all
$p<1$, response is bending dominated with $G \propto \kappa$;
at sufficiently large $\bmr$, response is stretching dominated
with $G\propto \mu$ as shown in the phase diagram of
Fig.~\ref{FIG:phasediagram}.

\section{Numerical Simulations}\label{SEC:numericalSimulations}

In the numerical portion of our work, we study the elasticity
of the filamentous kagome lattice by generating diluted lattice
conformations on a computer and then calculating their
mechanical response numerically. Practically this is done via
deforming the network by imposing a certain $\eta$ and by then
minimizing the elastic energy~(\ref{EQ:Hami}) over the
non-affine displacements $\delta \vu_\ell = \vu_\ell - \eta
\vr_\ell$ of the sites using a conjugate gradient algorithm. To
explore the response to shear, e.g., we set $\eta_{ij}= \gamma
(\delta_{ix} \delta_{jy} +  \delta_{iy} \delta_{jx})$, where
$\gamma$ specifies the magnitude of the applied deformation. We
use the same small magnitude $\gamma = 0.01$ for all
deformations. For a range of $p$-values, we generate up to $M
=200$ random conformations, we calculate several measurable
quantities for multiple $\kappa$-values, and we average
arithmetically over all conformations. The quantities that we
calculate are the elastic moduli and the corresponding
fractions of rigid conformations $n_{ijkl}$ and non-affinity
parameters $\Gamma_{ijkl}$. $n_{ijkl}$ is defined as the number
of conformations with non-zero $C_{ijkl}$ (non-zero meaning
larger than a small numerical threshold, here $10^{-8}$)
divided by $M$. The non-affinity parameters~\cite{DiDonna2005}
\begin{align}
\Gamma_{ijkl} = \frac{1}{N\gamma^2} \sum_\ell \left(  \delta \vu^0_\ell \right)^2,
\end{align}
where $N$ is the total number of sites and $\delta \vu^0_\ell$
is the equilibrium non-affine displacement of site $\ell$ in
the presence of the $\eta$ that leads to $C_{ijkl}$, measure
the deviation from a homogeneous strain field. To mitigate
boundary effects, we apply periodic boundary conditions on all
boundaries. We simulate system sizes ranging from $6^2$ to
$200^2$ unit cells which corresponds to $N=108$ and $N=120,000$
sites, respectively.

\begin{figure}
\includegraphics[width=.45\textwidth]{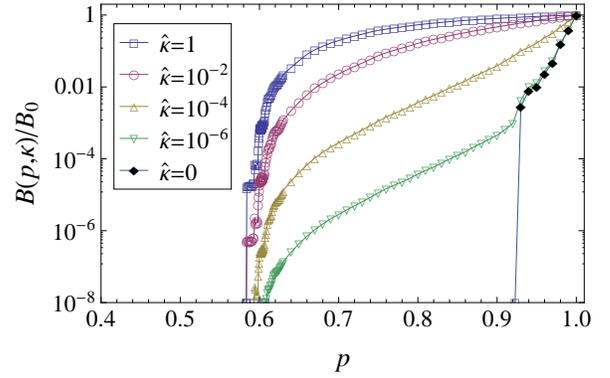}
\caption{The bulk modulus $B$ (normalized by the bulk
    modulus $B_0$ at $p=1$) from our simulations of $48^2$ unit cells.}
\label{FIG:BLogPlot}
\end{figure}

Figures~\ref{FIG:Compare} and \ref{FIG:BLogPlot} show
log-linear plots of the shear modulus $G$ and the bulk modulus
$B$, respectively, as functions of the occupation probability
$p$ for $48^2$ unit cells.  The EMT predicts the bulk modulus
to be proportional to $\mu_m$, and hence, both $\mu$ and $B$
should follow the same curves when they are normalized by their
respective affine values $G_0$ and $B_0$ at $p=1$. Within the
numerical errors, this prediction is indeed borne out by
Figs.~\ref{FIG:Compare} (a) and (b). Moreover, these figures
are consistent with the EMT prediction that the rigidity
percolation threshold $p_b$ be the same for all $\kappa > 0$.
Numerically, we find $p_b= 0.605 \pm 0.005$ which is only
slightly smaller than the EMT II prediction $p_b \approx 0.616$
and about 15\% smaller than the EMT I prediction $p_b \approx
0.692$. Previous EMT predictions for elastic networks have
produced somewhat larger values for the rigidity threshold than
found in corresponding numerical studies, see e.g.,
Refs.~\cite{Broedersz2010,Mao2011}, and this apparent trend is
continued here. For $\kappa=0$, the EMT predicts a first-order
rigidity percolation transition at $p=1$. The curves for
$\kappa=0$ in Figs.~\ref{FIG:Compare} (a) and (b) rise from
zero with finite slope below $p=1$, a clear finite-size effect.
Below, we will analyze this  finite-size effect systematically,
and we will find that our data is indeed consistent with a
first-order transition at $p=1$.

\begin{figure}
\includegraphics[width=8.4cm]{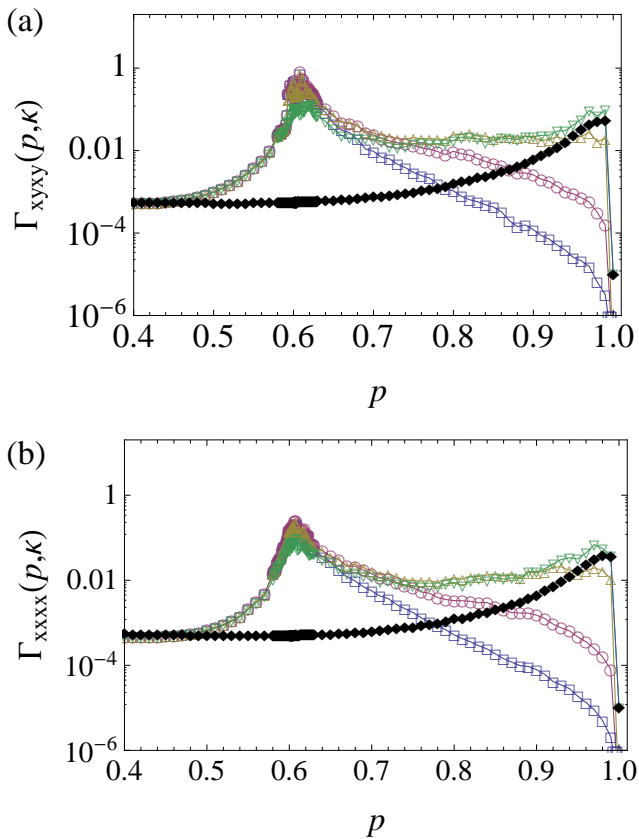}
\caption{(Color online) Non-affinity ratios (a) $\Gamma_{xyxy}$ and
(b) $\Gamma_{xxxx}$ for $48^2$ unit cells. Color-code is the same as in Figs.~\ref{FIG:Compare} and \ref{FIG:BLogPlot}.}
\label{fig:GammaCombined}
\end{figure}

Figure~\ref{fig:GammaCombined} shows the non-affinity functions
$\Gamma_{xyxy}$ and $\Gamma_{xxxx}$ extracted from the
simulation runs producing the results for $\mu$ and $B$
discussed in the previous paragraph. For $p=1$, the network is
affine, and the non-affinity functions are zero. The curves for
$\kappa >0$ are expected to have their maxima at $p_b$, and our
data is consistent with that expectation. The data for $\kappa
=0$ the curves for $G$ and $B$ should show no indication of the
existence of $p_b$, and indeed they do not.

\begin{figure}
\includegraphics[width=.45\textwidth]{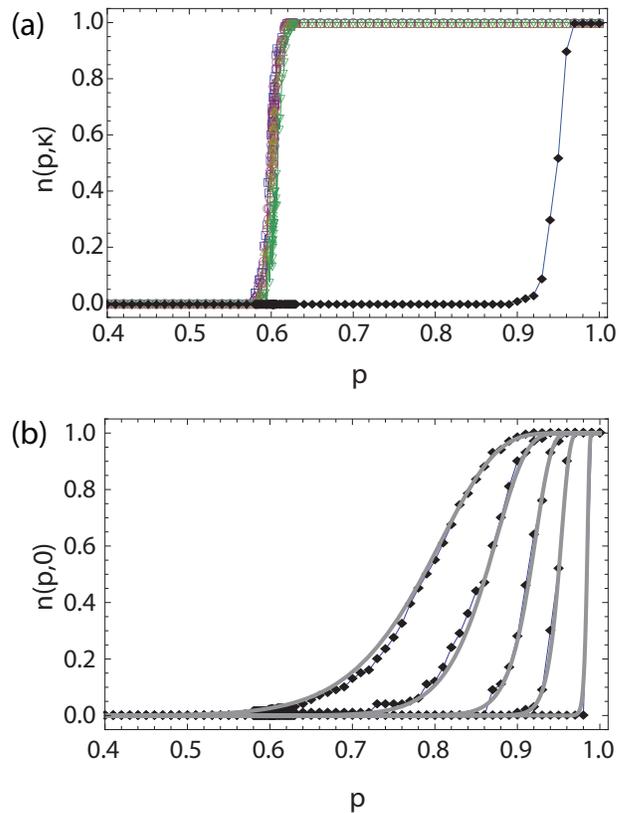}
\caption{(Color online) Fraction of rigid conformations for (a) $48^2$
unit cells and various values of $\kappa$ [with the same color-code as in Figs.~\ref{FIG:Compare} and \ref{FIG:BLogPlot}.], and (b) for $\kappa=0$ and
system sizes of $S^2$ unit cells with $S= 6, 12, 24, 48, 200$
(increasing from left to right). The bold lines correspond to the
theoretical result~(\ref{nMuTheotetical}).}
\label{fig:nMuCombined}
\end{figure}

Next, we turn to the fractions of rigid conformations. These
quantities are convenient for the purposes of finite-size
analysis, in particular for the expected first-order transition
for $\kappa =0$ at $p=1$. In the following, we will with focus
on $n = n_{xyxy}$. Figure~\ref{fig:nMuCombined}(a) displays $n$
for $48^2$ unit cells for $\kappa =0$ and several values of
$\kappa> 0$. The curves for $\kappa > 0$ are consistent with
the above statement that the rigidity percolation threshold for
all non-vanishing values of $\kappa$ is at $p_b= 0.605 \pm
0.005$.
Figure~\ref{fig:nMuCombined}(b) shows $n$ for $\kappa =0$ for a
variety of system sizes of $S^2$ unit cells. Along with the
data points, the figure follows predictions for $n(\kappa=0, p,
S)$ that follow from the elementary combinatorics (see
Sec.~\ref{SEC:elasticfil}). Consider a system such as that
shown in Fig.~\ref{FIG:diluted_kagome} with $N_x = 2S$ bonds
along its two sides.  Only those filaments, which consist of
$N_x$ bonds, that extend from the top to the bottom of the
sample will contribute to $G$, and $G$ will be zero unless at
least one of the filaments starting on bottom reaches the top,
which occurs with probability
\begin{align}
P_{\text{fil}}(p, N_x) = 1 - (1-p^{N_x})^{N_x}.
\end{align}
This function becomes a step function at p=1 when $N_x
\rightarrow \infty$.  Thus, we expect the data for $\kappa =0$
to be fit by the function
\begin{align}
\label{nMuTheotetical}
n (\kappa=0, p, S) = P_{\text{fil}}(p, 2S) .
\end{align}
Figure~\ref{fig:nMuCombined} (b) reveals that the data is fit
by this prediction remarkably well even though there is not a
single adjustable fit-parameter involved. Now, we are in
position to discuss the infinite-size limit. For $S\to \infty$,
$n (\kappa=0, p, S)$ approaches a unit-step function at $p=1$.
This establishes that the rigidity percolation transition for
$\kappa =0$ is a first-order transition at $p=1$ where the
shear modulus jumps discontinuously from zero to its affine
value.

Now, let us look at the elastic response near $p=1$.
Figure~\ref{fig:scalingNearP=1} shows the shear modulus and
$\mu_m$ as functions of the scaling variable $\tau$ defined in
Eq.~(\ref{EQ:tau}). We find remarkably good agreement between
the simulation data, the numerical solution of the EMT
equations as well as the EMT prediction for the asymptotic
scaling function $M (\tau)$ including the predicted value for
the constant $A$. Note that this agreement is obtained without
adjusting any fit-parameters.

\begin{figure}
\includegraphics[width=8.4cm]{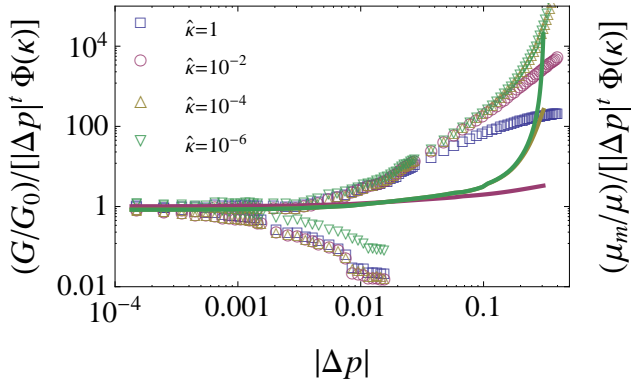}
\caption{(Color online) Scaling behavior of the shear modulus near $p_b$. The plot shows both our simulation results and our numerical solutions of the EMT I equations relative to the asymptotic scaling form near $p_b$ predicted by the EMT, cf.\ Eq.~(\ref{EQ:scaling-2}).
The data points stem from our simulations for $48^2$ unit cells, and we
use $p_b = 0.605$, $c_1 = 0.3$, $c_2 = 0.007$, and $t=0.2$. The lower branches
correspond to values of $p$ below $p_b$ and are a typical finite-size effect.
The continuous curves represent the numerical solutions to the EMT I equations, which reduce to Eqs.~\eqref{EQ:AppBAsym1} to\eqref{EQ:AppBAsym2} as $\Delta p$ tends to zero with $c_1 = 0.03802$, $c_2 = 4.697$, and $p_b = 0.692$.}
\label{fig:scalingNearPb}
\end{figure}

Figure~\ref{fig:scalingNearPb}, finally, displays the scaling
behavior near the rigidity percolation threshold $p_b$. Our
numerical data collapses in full qualitative agreement with the
EMT scaling form~(\ref{EQ:scaling-2}) albeit with an exponent
$t=0.2$ that is significantly smaller than the $t=1$ predicted
by the EMT. The value $c_1 = 0.25$, on the other hand, is in
semi-quantitative agreement with the EMT.

\section{Discussion}\label{SEC:discussion}

We have introduced the diluted kagome lattice in
$2$-dimensions as a model for networks of semi-flexible
polymers, whose maximum coordination number is four.  We
identify straight lines of contiguous occupied bonds as
filaments that, following previous work
\cite{Wilhelm2003,Head2003,Head2003a,Heussinger2006,HeussingerFre2007a,HuismanBar2008},
we endow with the energy of an elastic beam characterized by a
stretching (Young's) modulus $\mu$ and a bending modulus
$\kappa$.  We contrast this model, which has filaments of
arbitrary length, with the Mikado lattice whose filaments are
of finite (and usually fixed length).  We show that when
$\kappa = 0$, this kagome model has affine response and
non-vanishing elastic moduli so long as there are
sample-traversing filaments, which is the case for samples with
finite lengths and widths $W$, even for bond-occupation
probabilities $p$ less than one.  As $W$ increases, the
probability of sample-traversing filaments decreases, and in
the $W\rightarrow \infty$ limit, elastic moduli, which are
nonzero for undiluted $p=1$ lattice, fall precipitously to zero
for $p=1^-$ in a first-order transition. The undiluted lattice
has $\kappa$-independent, and thus affine, elastic moduli. The
addition of bending forces restores rigidity for any $\kappa$
for $p$ greater than $p_b$, the rigidity percolation threshold,
and elastic moduli approach the non-zero affine values of the
$p=1$ lattice as $p\rightarrow 1$. We argue that this is the
underlying cause for the affine limit found in the large
$l/l_c$ limit ($p$ near one) in the Mikado model.

We use two recently introduced lattice-based effective medium
theories to calculate elastic moduli of the diluted kagome
lattice as a function of $\mu$, $\kappa$, and $p$, we we
calculate scaling forms for the shear modulus both near $p=1$
and $p=p_b$.  Both forms are a function of the unitless ratio
$\bmr=\kappa/(\mu a^2)$, where $a$ is the lattice spacing, and
yield a crossover for all $p_b<p<1$ between bending dominated
response at small $\bmr$ and stretching dominated response at
large $\bmr$. We supplement our EMTs with numerical simulations
of the shear modulus and other functions, such as that
measuring the degree of nonaffine response.  The results of
these simulations agree qualitatively with EMT predictions for
all $p$ and quantitatively with them near $p=1$.

Our study of the kagome lattice provides insight into the
behavior of $3$-dimensional filamentous lattices with maximum
coordination number of four.  With stretching forces only,
these lattices are subisostatic with of order one zero mode per
site, and one might, therefore, conjecture that their elastic
moduli should vanish when $\kappa = 0$.  However, simulations
\cite{Stenull2011,BroederszMac2012} on two model lattices
consisting of straight filaments yields curves of elastic
moudli as a function of $p$ that look very similar to those of
the kagome lattice presented here with an approach to affine
$\kappa$-independent response as $p\rightarrow 1$. The
underlying cause of this result is that straight
sample-traversing filaments in \emph{any} dimension give rise
to nonvanishing macroscopic elastic moduli.  Indeed, exact
calculations \cite{Stenull2011} for one of the $3d$-lattices
with elastic-beam energies show that all elastic moduli are
nonzero when $\kappa = 0$ and $p=1$ and that response is
affine. Near $p=1$, simulations on both of the $3d$ lattice
show a crossover from a bending dominated regime with $G\sim
(\kappa/a^3)(L/a)^2$ to a stretching dominated regime in which
$G \sim \mu$ in agreement with the predictions of our
kagome-based lattice EMT. In fact the results of one of the
simulations \cite{Stenull2011} track the scaling function of
the our kagome EMT. We conjecture that the important common
feature of the $2d$ and $3d$ lattices is that they both consist
of sample-traversing straight filaments in the undiluted limit.
Away from $p=1$, bending forces stabilize the lattices in a
process that is not so sensitive to lattice dimension, even
though the central-force $3d$ lattice is subisostatic.  Our
EMTs provide some support for this conjecture.  The
characteristic property of the kagome EMT solutions, such as
the form of the scaling function near $p=1$, are a direct
consequence of the existence of lines of zeros in the phonon
spectrum at $\kappa=0$ that get raised to nonzero frequency for
$\kappa >0$ as shown in Fig.~\ref{FIG:Integral}. This feature
causes the integral $H_{2,s}$ [Eqs.~(\ref{EQ:H2pi2}) to
(\ref{EQ:H23final})] to diverge as $\sqrt{\bmr}$ as $\bmr =
\kappa/(\mu a^2) \rightarrow 0$. In three dimensions, the lines
of zero modes in the phonon spectrum become planes of
zero-modes, and we believe that the $3d$ version of $H_{2,s}$
will have a from analogous to Eq.~(\ref{EQ:H23final}) but with
$g(q_y)$ replaced by a similar function of $q_y$ and $q_z$:
\begin{align}
	H_{2,s}(\bmr,0)&\sim \int d^3 q \frac{\vert \langle \vB_{1}^{b} \vert \nu_f\rangle\vert^2}
	{c q_x^2 + \bmr  \tilde{g} (q_y,q_z)} \nonumber\\
	&\sim \bmr^{-1/2} .
\end{align}
When $b_m=0$, the denominator vanishes for all $q_y$ and $q_z$
when $q_c = 0$, and the integral diverges.  Nonzero $b_m$
yields a $b_m^{-1/2}$ divergence just as in the $2d$ kagome
case. Unfortunately, the $3d$ lattices are quite complicated,
with a $54$-site unit cell in one case \cite{Stenull2011} and
complicated crossing configurations in the other
\cite{BroederszMac2012}, and we have not been able to set up a
$3d$ EMT to verify this conjecture.

Much of the work on the elastic properties of filamentous
networks (with the notable exceptions of
\cite{HuismanBar2008,HuismanBar2010,HuismanLub2011}) have
focused on models, such as the Mikado or kagome lattice
presented her, consisting of straight filaments. This work has
also largely focused on what are really mechanical models with
beam energies assigned to the filaments and the effects of
thermal fluctuations ignored. As our analysis shows, these are
exceptional lattices because they can support stress with
central forces only even when their coordination number $z$  is
substantially less the Maxwell stability limit of $z=2d$.
Filaments in real biological gels are not straight, and their
energies are not described by that of an elastic beam. It is
thus legitimate to ask how seriously one should take models
such as those presented here as realistic descriptions of real
bio-gels of semiflexible polymers.  If constituent filaments
are not straight, then in general, at least some of elastic
moduli of even perfect lattices vanish in the absence of
bending forces.  Prime examples of this behavior are found in
the two-dimensional honeycomb lattice and the three-dimensional diamond lattice
whose bulk moduli are nonzero but whose shear moduli vanish in
the  $\kappa\rightarrow 0$ limit, but it has also been observed
in a more realistic model of a filamentous lattice
\cite{HuismanLub2011}. If $\kappa$ is needed to stabilize the
system, the elastic response to external stresses will involve
bending and thus be nonaffine. So, it would seem that the
straight-filament models are not such good ones for real
systems, though they do provide us with valuable insight into
how network architecture influences elastic response.  One of
the predictions of the straight-filament models is that
response becomes more stretching dominated and more affine as
$\kappa$ increases. Thus, it is plausible that if $\kappa$ is
large enough, the elastic response of even those lattices whose
shear moduli vanish if $\kappa = 0$ will exhibit stretching
dominated, nearly affine response for sufficiently large
$\kappa$.

\begin{acknowledgments}
We are grateful for helpful discussions with Fred MacKintosh
and Chase Broedersz.  This work was supported in part by
National Science Foundation under DMR-1104707 and under the
Materials Research Science and Engineering Center DMR11-20901 .
\end{acknowledgments}

\appendix

\section{EMT Generalities and the EM Dynamical
Matrix}\label{SEC:APP1}

To implement the $T$-matrix version of EMT, we deal with the
dynamical matrix $\DMm$ of the EM, its associated Green's
funcition $\GMm= -(\DMm)^{-1}$ (we consider only zero
frequency), the perturbation $\PPV$ associated with bond
replacement, the dynamical matrix $\DM = \DMm + \PPV$, and its
associated Green's function $\GM= - \DM^{-1}$.  [For more
details of this formalism, see Refs.~\cite{Mao2011a} and
\cite{Mao2011}] Because there are three sites per unit cell,
all of these matrices are $6N \times 6N$ matrices, to be
detailed further below, where $N$ is the number of unit cells
in the lattice. We will specify $\GMm$ more completely below.
With these definitions,
\begin{equation}
\GM = \GMm + \GMm\cdot \TM \cdot \GMm = (\GMm-\PPV)^{-1} ,
\end{equation}
where
\begin{equation}
\TM = \PPV\cdot(\IM - \GMm\cdot \PPV)^{-1} .
\end{equation}
The EMT  self-consistency equation requires that the disorder
average over the probability distribution of Eq.~(\ref{EQ:P-I})
for EMT I or Eq.~(\ref{EQ:P-II}) or EMT II of the perturbation
vanishes
\begin{equation}\label{EQ:SCET}
   \langle \TM (\ss,\bs )\rangle =0 .
\end{equation}

All of the matrices discussed here can be expressed in a
position or wavenumber representation.  The six independent
displacements in a unit cell can be expressed as a
six-dimensional vector $\U_\vl = (\vu_{\vl,1}, \vu_{\vl,2},
\vu_{\vl,3})$ for each $\vl$ or as its Fourier transform
$\U_\vq =\frac{1}{N} \sum_\vl
e^{-i\vq\cdot\vr_\ell}\U(\vl)=(\vu_{\vq,1},\vu_{\vq,2},\vu_{\vq,3})$.
The energy of the bond-replaced system is then
\begin{eqnarray}
E & = &\frac{1}{2} \sum_{\vl,\vl'}\U_\vl \DM_{\vl,\vl'} \U_\vl'
\nonumber \\
& =& \frac{1}{2N^2} \sum_{\vq,\vq'} \U_{-\vq} \DM_{\vq,\vq'}
\U_{\vq'} ,
\end{eqnarray}
where $D_{\vl,\vl'}$ and $\DM_{\vq,\vq'}$ are $6 \times 6$
dimensional matrices for each pair $(\vl,\vl')$ and
$(\vq,\vq')$.  The effective medium is translationally
invariant and
\begin{equation}
\DMm_{\vq,\vq'} = N \delta_{\vq,\vq'} \DMm_{\vq} .
\end{equation}

The energy of the effective medium is constructed by occupying
all bonds with identical beams with stretching and bending
moduli $\mu_m$ and $\kappa_m$ and adding (for EMT I) an
additional coupling of strength $\lambda_m$ coupling angles in
neighboring $NNN$ bonds along a filaments [See
Ref.~\cite{Mao2011} for details]. The EM energy is then
\begin{eqnarray}
E^m & = & E_s(\mu_m) + E_b(\kappa_m) + E^m_{bb}(\lambda_m) \\
&\equiv&\frac{1}{2N}\sum_{\vq}\U_{\vq}\cdot \DMm_{\vq} \cdot \U_{-\vq}
\end{eqnarray}
where $E_s(\mu_m)$ and $E_b(\kappa_m)$ are evaluated at
$g_{\ell,\ell'}=1$ for all bonds $\langle \ell,\ell'\rangle$,
where
\begin{equation}\label{EQ:Eeff_bb}
E^m_{bb}(\lambda_m) = \frac{\lambda_m}{\a^3} \sum_{\ell_2,\ell_3}
\theta^n_{\ell_1,\ell_2,\ell_3} \theta^n_{\ell_2,\ell_3,\ell_4} ,
\end{equation}
where it is understood that $\ell_1,\ell_2,\ell_3$ and
$\ell_4$ are all contiguous sites along a single filament.
$\GMm_{\vq}$ can be constructed from the Fourier transforms of
the stretching and bending energies on $NN$ and phantom $NNN$
bonds.  For example, the stretching energy of bond $5$ in
Fig.~\ref{fig:simpleKagome}, connecting site $\ell=(\vl,3)$ at
position $\vr_\vl -\uvepara_3$ with site $\ell'=(\vl,2)$ at
position $\vr_\vl + 2 \uvepara_2+\uvepara_1$ such that
$\vr_\ell'-\vr_\ell= \uvepara_2$ is
$\frac{1}{2}(\mu_m/\a)[(\uv_{\ell'}-\uv_\ell)\cdot\uvepara_2]^2$.
The Fourier transform of
$(\uv_{\ell'}-\uv_\ell)\cdot\uvepara_2$ is
\begin{equation}
e^{-i\vq\cdot\vl}(e^{-i \vq\cdot
\ev_2}\vu_{2,\vq}-\vu_{3,\vq})\cdot \uvepara_2 =e^{-i\vq\cdot \vr_{\vl}}
\vB^2_{5,\vq}\cdot\U_\vq ,
\end{equation}
where $\vB^2_{5,\vq}$ is specified in detail in
Eq.~(\ref{EQ:Bs}) below.  Similar procedures apply to all
stretching and bending bonds, and the EM dynamical matrix can
be expressed as
\begin{eqnarray}\label{EQ:DMCons}
    &&\DMm_{\vq}\left(\sm,\bm,\lm \right)  \nonumber\\
    &=& \frac{\sm}{a} \sum_{n=1}^{6} \vB^{s}_{n,\vq} \vB^{s}_{n,-\vq} \nonumber\\
         &&+ \frac{\bm}{a^3} \sum_{n=1}^{6} \vB^{b}_{n,\vq} \vB^{b}_{n,-\vq}
	\nonumber\\ &&
    + \frac{\lm}{a^3} \sum_{n=1}^{6}
    2\cos(\vq\cdot\uvepara_{n}) \vB^{b}_{n,\vq} \vB^{b}_{n,-\vq} ,
\end{eqnarray}
where
\begin{align}\label{EQ:Bs}
	\vB_{1,\vq}^{s} &= \{-\uvepara_{1},\uvepara_{1},0,0  \}
	\nonumber\\
	\vB_{2,\vq}^{s} &= \{0,0,-\uvepara_{2},\uvepara_{2} \}
	\nonumber\\
	\vB_{3,\vq}^{s} &= \{\uvepara_{3},0,0,-\uvepara_{3} \}
	\nonumber\\
	\vB_{4,\vq}^{s} &= \{e^{-i \vq\cdot \uvepara_{1}}\uvepara_{1},
		-\uvepara_{1},0,0  \}
	\nonumber\\
	\vB_{5,\vq}^{s} &= \{0,0,e^{-i \vq\cdot \uvepara_{2}}\uvepara_{2},
		-\uvepara_2 \}
	\nonumber\\
	\vB_{6,\vq}^{s} &= \{-\uvepara_{3},0,0,e^{-i \vq\cdot \uvepara_{3}}\uvepara_{3} \}
\end{align}
and
\begin{align}
\vB_{1,\vq}^b & = 2\{2 \uveperp_1, -(1+e^{i\vq\cdot\uvepara_1})\uveperp_1,0,0\} ,\nonumber\\
\vB_{2,\vq}^b &=2 \{0,0, 2 \uveperp_2,-(1+e^{i\vq\cdot\uvepara_2})\uveperp_2\}  , \nonumber\\
\vB_{3,\vq}^b &= 2\{-(1+e^{i\vq\cdot\uvepara_3})\uveperp_3,0,0, 2 \uveperp_3)\}  , \nonumber\\
\vB_{4,\vq}^b & =2 \{-(1+e^{-i \vq\cdot \uvepara_1})\uvepara_1, -2\uvepara_1,0,0\} ,\nonumber\\
\vB_{5,\vq}^b &= 2\{0,0,-(1+e^{-i\vq\cdot\uvepara_2})\uveperp_2, 2 \uveperp_2)\}, \nonumber \\
\vB_{6,\vq}^b &= 2 \{2 \uveperp_3,0,0, -(1+e^{-i\vq\cdot\uvepara_3})\uveperp_3)\} ,
\end{align}
where $\uveperp_{n}$ are the unit vectors perpendicular to the
bonds, $\uvepara_{n}$ should be understood as
$\{\epara_{n,x},\epara_{n,y}\}$, and the same for
$\uveperp_{n}$ so all these vectors are $6$ dimensional.  The
factor of $2$ is from the fact that the length of the bonds is
$1/2$.

The scattering $\mathbf{V}$ can also be expressed in this form.
We assume that the changed bond is bond $1$ in the unit cell at
$\vec{r}=0$ as marked in Fig.~\ref{fig:simpleKagome}.  Then we
have
\begin{eqnarray}\label{EQ:VEMT}
    \mathbf{V}_{\vq,\vq'}
	&=& \frac{\mu_s-\sm}{a} \vB^{s}_{1,\vq} \vB^{s}_{1,-\vq} \nonumber\\
         &&+ \frac{\bs-\bm}{a^3}  \left(\vB^{b}_{1,\vq} \vB^{b}_{1,-\vq}
         	+ \vB^{b}_{4,\vq} \vB^{b}_{4,-\vq}\right)
	\nonumber\\ &&
    - \frac{\lm}{a^3}
    \left(\vB^{b}_{1,\vq} \vB^{b}_{4,-\vq} +\vB^{b}_{4,\vq} \vB^{b}_{1,-\vq} \right)
\end{eqnarray}

\section{Asymptotic solutions of the EMT I equations}\label{APP:Asym}
Asymptotic solutions for the EMT I self-consistency
equation~\eqref{EQ:SCET} has been developed in
Ref.~\cite{Mao2011} in the limit of small $\b/\s$.   In this
section we review this asymptotic solution and apply it to the
case of kagome lattice. For convenience we define the notation
\begin{align}
	\bmr &\equiv \bm/(\sm a^2) \nonumber\\
	\lmr &\equiv \lm/(\sm a^2) \nonumber\\
	b &\equiv \b/(\sm a^2) \nonumber\\
\end{align}
and 
\begin{align}
	\mathbf{H} \left(\bmr,\lmr\right) &\equiv -\frac{\sm}{a}
		\GF \left( \sm,\bm,\lm\right) \nonumber\\
		&=- \GF \left( 1,\bmr,\lmr \right) ,
\end{align}
where the second line is derived from the definition of the
dynamical matrix. The self-consistency equation~\eqref{EQ:SCET}
can be solved by projection to the space $\vert
\vB_{1}^{s}\rangle$, $\vert \vB_{1}^{b}\rangle$, $\vert
\vB_{4}^{b}\rangle$ that spans $\mathbf{V}$.  In this basis we
can rewrite the EMT matrix equation into three independent
equations
\begin{widetext}
\begin{subequations}
\begin{align}
	&\sm = \s \frac{\Prob - H_{1} \left(\bmr,\lmr\right) }
		{1 - H_{1} \left(\bmr,\lmr\right)}
		\label{EQ:AsymEqOne} \\
	&2\Big(\frac{1}{\br}+\frac{1}{\bmr}\Big)^{-1}
    \Big(\Prob-\frac{1}{2}\big(1+\frac{\bmr}{\br}\big) - \bmr H_{2}-\lmr H_{3} \Big)
     +(\bmr^2+\lmr^2)H_2 + 2\bmr\lmr H_{3} =0, \label{EQ:AsymEqTwo}\\
    & -\lmr - 2\Big(\frac{1}{\br}+\frac{1}{\bmr}\Big)^{-1} \big(\lmr H_{2}+\bmr H_{3} \big)
     +2\bmr\lmr H_2 + (\bmr^2+\lmr^2)H_{3} =0 \label{EQ:AsymEqThree},
\end{align}
\end{subequations}
\end{widetext}
where $H_1$ and $H_3$ are the projections of the Green's
function defined as
\begin{align}
	H_1(b,l)&\equiv \langle \vB_{1}^{s} \vert
	\mathbf{H}(b,l) \vert \vB_{1}^{s}\rangle \nonumber\\
	H_2(b,l)&\equiv \langle \vB_{1}^{b} \vert
	\mathbf{H}(b,l) \vert \vB_{1}^{b}\rangle \nonumber\\
	H_3(b,l)&\equiv \langle \vB_{1}^{b} \vert
	\mathbf{H}(b,l) \vert \vB_{4}^{b}\rangle ,
\end{align}
with the inner product defined as
\begin{align}
	&\langle \vB_{1}^{s} \vert \mathbf{H}(b,l) \vert \vB_{1}^{s}\rangle  \nonumber\\
	=& \frac{1}{N}\sum_{\vq}
		\vB_{1,-\vq}^{s} \cdot \mathbf{H}(b,l) \cdot \vB_{1,\vq}^{s} \nonumber\\
     = &\textrm{Tr} \mathbf{H}(b,l)[ \vert \vB_{1}^{s}\rangle \langle \vB_{1}^{s}\vert ]
\end{align}
where the sum is over all $N$ vectors in the first Brillouin
zone of the kagome lattice, and the trace is understood to
include the sum over these vectors (along with the factor of
$1/N$ in addition to the sum over the $6$ dimensional space of
$\vB_{1,-\vq}^{s}$. These equations are exactly equivalent to
the matrix equation~\eqref{EQ:SCET}.

For the special case of $\b=0$, equations~(\ref{EQ:AsymEqOne},
\ref{EQ:AsymEqTwo}, \ref{EQ:AsymEqThree}) simplify and give
\begin{align}\label{EQ:zeroS}
	&\sm = \s \frac{\Prob - H_{1} \left(0,0 \right) }
		{1 - H_{1} \left(0,0 \right)}\\
	&\bmr=0 \\
	&\lmr=0
\end{align}
As discussed in Ref.~\cite{Mao2011}, it is straightforward
from the definition  $\DM$  and the fact that the central force
undiluted kagome lattice is isostatic with $z=2d$ to derive the
relation
\begin{align}
	H_1(0,0)=1,
\end{align}
The effective medium filament stretching stiffness is then
$\sm=0$ for $\Prob<1$ and $\sm=\s$ for $\Prob=1$. Therefore
this $\b=0$ EMT solution indicates a first order rigidity
transition at $\Prob=1$.

In the following we solve these equations asymptotically at
small $\b$ near the two critical points $\Prob=1$ and $\Pb$.

\subsection{Asymptotic solution near $\Prob=1$}\label{APP:Asym1}
The EMT solution for small $\b>0$ can be calculated
perturbatively from the $\b=0$ solution~\eqref{EQ:zeroS}.  Near
$\Prob=1$, as discussed in Ref.~\cite{Mao2011}, we can make
simplifications to the self-consistency
equation~(\ref{EQ:AsymEqTwo},\ref{EQ:AsymEqThree}) using the
fact that $\b\ll 1$ so that
\begin{align}
	\bm &= \b (2\Prob-1) \nonumber\\
	\lm &= 0,
\end{align}
(which we shall verify later) and therefore we only need to
solve Eq.~\eqref{EQ:AsymEqOne} using perturbation.

In contrast to the perturbative calculation in the triangular
lattice, the kagome lattice effective medium is isostatic as
$\bm\to 0$, and thus the phonon Green's functions exhibit
singularities.  These singularities correspond to the
zero-frequency floppy modes of the dynamical matrix, and make
diverging contributions to $\mathbf{H}$.

Nevertheless, perturbation theory around the $\b=0$ solution is
still well defined. We shall see below that the $\mathbf{H}$
singularity is proportional to $\bmr^{-1/2}$ and thus all the
terms in the self-consistency equations are non-singular.  The
expansion of $H_1$ at small $\bmr$ can be calculated using the
following equality
\begin{align}
	&\langle \vB_{1}^{s} \vert \mathbf{H}(\bmr,0) \vert \vB_{1}^{s}\rangle
	+\bmr \langle \vB_{1}^{b} \vert \mathbf{H}(\bmr,0) \vert \vB_{1}^{b}\rangle \nonumber\\
    & =  \textrm{Tr}\mathbf{H}(\bmr,0)(\vert \vB_{1}^{s}\rangle\langle \vB_{1}^{s} \vert 
    +\bmr\vert \vB_{1}^{b}\rangle\langle \vB_{1}^{b} \vert ) \nonumber\\
    & =  \frac{1}{6} \textrm{Tr}\mathbf{H}(\bmr,0)
    \left[\sum_{n=1}^6(\vert \vB_{n}^{s}\rangle\langle \vB_{n}^{s} \vert 
    +\bmr\vert \vB_{n}^{b}\rangle\langle \vB_{n}^{b} \vert )\right] =1 .
\end{align}
These relations follow because all of the $6$ $NN$ and the $6$
$NNN$ bonds in a cell are, respectively, equivalent by symmetry
allowing the trace over one set of bonds to be replaced by
$1/6$ the trace over the sum of the bonds.  But the quantity in
square brackets is just the inverse of $\mathbf{H}(\bmr,0)$,
and the final result follows.

\begin{figure}
\centering
  \subfigure[]{\includegraphics[width=0.32\textwidth]{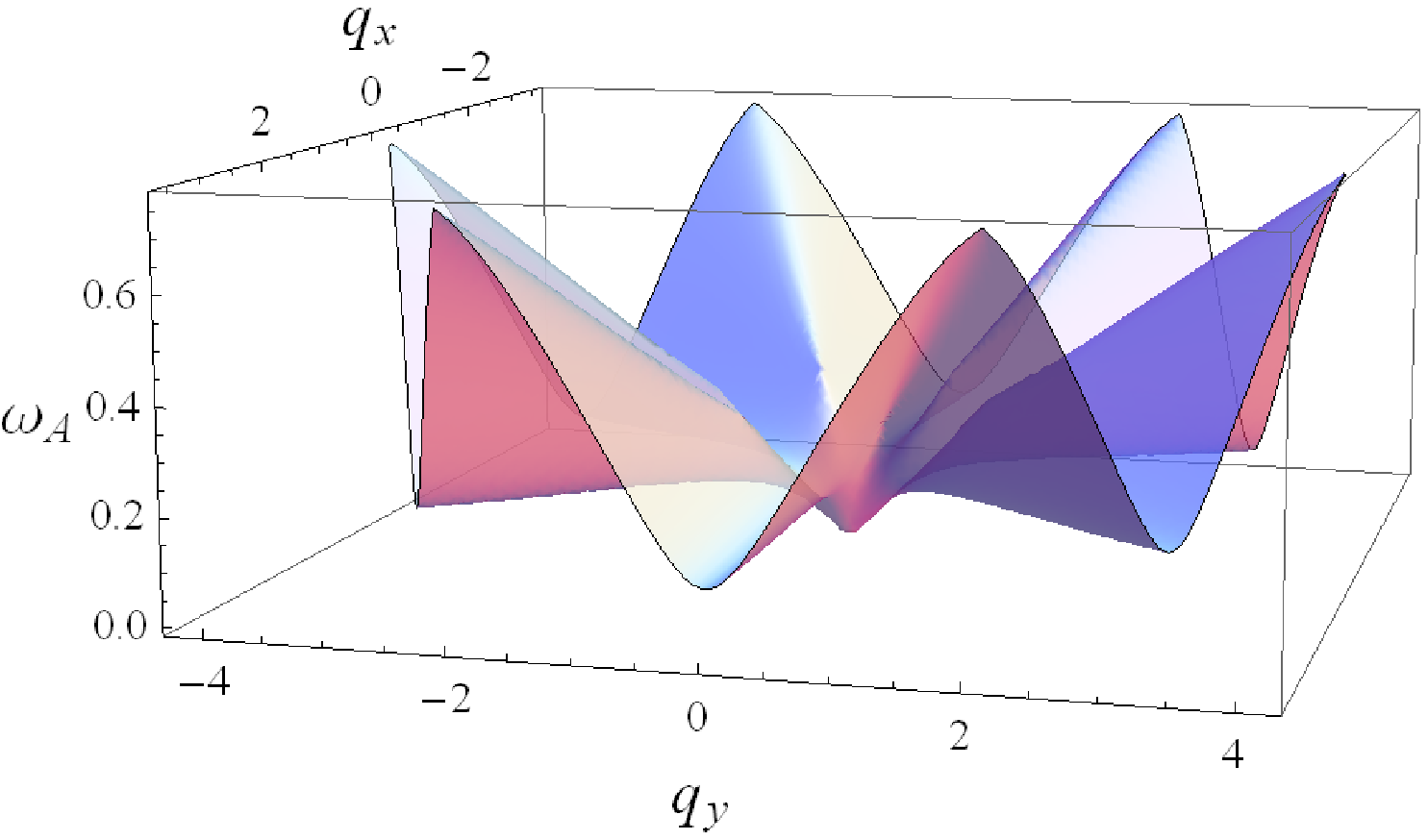}}
  \subfigure[]{\includegraphics[width =0.32\textwidth]{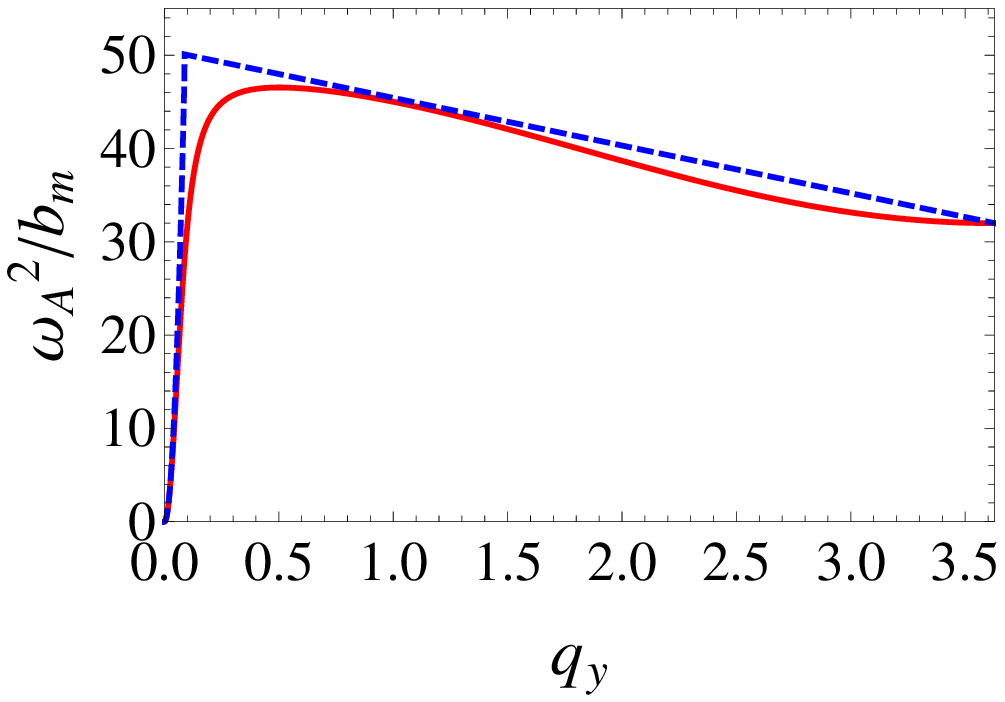}}
  \subfigure[]{\includegraphics[width =0.32\textwidth]{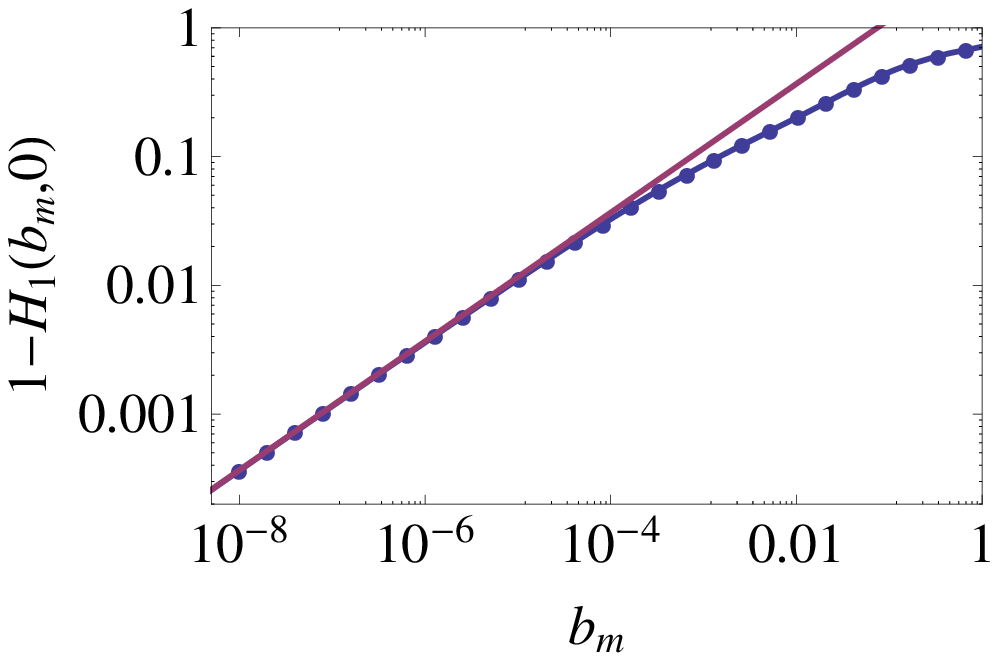}}
  \caption{(a) Dispersion relation of the floppy mode taking $\sm=1,\bm=0.0005,\lm=0$,
  where the frequency $\omega_A$ is the square root of the lowest eigenvalue of
  $\DM(\sm,\bm\lm)$.  (b) The floppy mode branch of the eigenvalues of the
  dynamical matrix along the direction of $\pi/2$ (which is $q_x=0$) taking
  $\sm=1,\bm=10^{-5},\lm=0$.  The red solid curve shows the actual eigenvalues
  calculated from the dynamical matrix numerically, and the blue
  dashed line shows the approximation we used in Eq.~(\ref{EQ:GFExp})\,
  that $\omega_A^2 = q_y^2/16  $ if $q_y<Q_s$ and $\omega_A^2 =16\bmr g(q_y)  $
  otherwise (only the $q_y>Q_s$ part is included in Eq.~\ref{EQ:GFExp}
  because other contributions are much smaller).
  (c) Comparison between numerical (blue points connected by line)
  and the asymptotic form~(\ref{EQ:Hone}) (red line) of $1-H_1(\bmr,0)$.}\label{FIG:Integral}
\end{figure}

Employing the analysis of the phonon modes in
Ref.~\cite{Mao2011a}, we find that $\vert \vB_{1}^{b}\rangle$
has a nonzero projection onto the floppy mode branch, which we
call $\vert \nu_f \rangle$, whereas $\vert \vB_{1}^{s}\rangle$
does not.  The floppy modes branch has low frequencies that are
proportional to $\sqrt{\bmr}$ along symmetry directions at
$\{\frac{\pi}{6},\frac{\pi}{2},\frac{5\pi}{6}
,\frac{7\pi}{6},\frac{3\pi}{2},\frac{11\pi}{6}\}$ in the first
Brillouin zone, as shown in Fig.~\ref{FIG:Integral}.  In this
calculation we shall use the direction $\pi/2$ which correspond
to $q_x=0$ as an example.  Near the $q_x=0$ line the floppy
modes branch phonon Green's function takes the form,
\begin{align}\label{EQ:GFExp}
	G_{\textrm{floppy},\frac{\pi}{2},\vq} = -\frac{1}{\frac{3}{16}q_x^2+16\bmr g(q_y)} ,
\end{align}
where
\begin{align}
	g(q_y) = \frac{3Q_M-2Q_S-q_y }{Q_M-Q_S}
\end{align}
with $Q_M=2\pi/\sqrt{3}$ and $Q_S=4(3\bmr/2)^{1/4}$ and thus
$g(q_y)$ takes value between $2$ and $3$ in the first Brillouin
zone.  This form of the Green's function is shown in
Fig.~\ref{FIG:Integral}. It was derived in Ref.~\cite{Mao2011a}
in which the weak additional interactions are NNN bonds rather
than bending forces.

It is straightforward to calculate the singular part of $
\langle \vB_{1}^{b} \vert \mathbf{H}(\bmr,0) \vert
\vB_{1}^{b}\rangle$ along symmetry directions of the floppy
modes, which involves the integral
\begin{align}\label{EQ:H2pi2}
	&H_{2,\frac{\pi}{2},s} (\bmr,0)\nonumber\\
	=&v_0 \int_{1BZ, \vert q_x\vert\le \vert q_y\vert/\sqrt{3}}
	\frac{d^2 q}{(2\pi)^2}
	\frac{\vert \langle \vB_{1,\vq}^{b}
	\vert \nu_f\rangle\vert^2 }{\frac{3}{16}q_x^2+\bmr g(q_y)}
\end{align}
where $v_0=\sqrt{3}/2$ is the area of the kagome lattice unit
cell, and the condition $\vert q_x\vert\le \vert
q_y\vert/\sqrt{3}$ confines the integral to be around the
floppy mode directions $\pi/2$ and $3\pi/2$ near which the
expansion~\eqref{EQ:GFExp} is valid.
The total contribution involves the singular part of all the
isostatic directions
\begin{align}\label{EQ:H23}
	H_{2,s} = H_{2,\frac{\pi}{2},s}+H_{2,\frac{\pi}{6},s}+H_{2,\frac{5\pi}{6},s}.
\end{align}
All these terms exhibit similar behavior so we just discuss
$H_{2,\frac{\pi}{2},s}$ for convenience.  In
Eq.~\eqref{EQ:H2pi2}, assuming $\bmr\ll1$, one can first
integrate out $q_x$ using contour integrals, which gives
\begin{align}
	H_{2,\frac{\pi}{2},s}(\bmr,0)& \simeq
		\frac{1}{(2\pi)^2 /v_0}\int_{0}^{\frac{2\pi}{\sqrt{3}}} dq_y
	\frac{4\pi }{\sqrt{3}}
	\frac{\vert \langle \vB_{1,\vq}^{b}
	 \vert \nu_f\rangle\vert^2 }{\sqrt{\bmr g(q_y)}} \nonumber\\
	& \sim \bmr^{-1/2}
\end{align}
indicating a leading order divergence $\bmr^{-1/2}$ of $H_2$ at
small $\bm$.  A calculation including all terms in
Eq.~\eqref{EQ:H23} yields
\begin{align}\label{EQ:H23final}
	H_{2,s} (\bmr,0)\simeq \frac{\sqrt{A}}{2}\bmr^{-1/2} ,
\end{align}
where
\begin{align}
	 \frac{\sqrt{A}}{2} =20(1-\sqrt{2/3}).
\end{align}
From this we get
\begin{align}\label{EQ:Hone}
	H_{1}(\bmr,0) \simeq 1-\frac{\sqrt{A}}{2}\bmr^{1/2}
\end{align}
at small $\bm$.  Plugging this back into
Eq.~\eqref{EQ:AsymEqOne}, we arrive at the solution
\begin{align}\label{EQ:mumAsym}
	\frac{\sm}{\s} =& \frac{(\Delta\Prob)^2}{A(2\Prob-1) \b/(\s a^2)} \nonumber\\
	& \cdot \left\lbrack -1+
	\sqrt{1+\frac{A(2\Prob-1) \b/(\s a^2)}{(\Delta\Prob)^2} } \right\rbrack^2
\end{align}
where $\Delta\Prob\equiv 1-\Delta \Prob$ and this solution is
asymptotically accurate for $\b\ll 1$ and $\Delta \Prob\ll 1$.

\subsection{Asymptotic solution near $\Prob=\Pb$}\label{APP:Asym2}
The rigidity threshold $\Pb$ can be obtained by plugging
$\sm=0$ into
Eqs.~(\ref{EQ:AsymEqOne},\ref{EQ:AsymEqTwo},\ref{EQ:AsymEqThree}),
which leads to
\begin{align}
	\Pb&\simeq 0.6920 , \nonumber\\
	\bmr &\simeq 0.03788, \nonumber\\
	\lmr &\simeq 0.008032.
\end{align}
Following the calculation in Ref.~\cite{Mao2011} we arrive at
the asymptotic solution for $\sm$
\begin{align}\label{EQ:AppBAsym1}
	\mu_m = \mu \Phi(\kappa/(\mu a^2))
	(\Prob-\Prob_b)
\end{align}
where
\begin{align}
	\Phi(x)=\frac{c_2 x}{c_1+x }
\label{EQ:scalefn-2}
\end{align}
with
\begin{align}\label{EQ:AppBAsym2}
	 c_1&\simeq 0.03802\nonumber\\
	 c_2 &\simeq 4.697.
\end{align}

\section{Asymptotic solutions to EMT II
equations}\label{SEC:AppC}

The self-consistency equation of the EMT II can be written
as~\cite{Das2007,Das2012}
\begin{align}
	\frac{\sm}{\s} &= \frac{p-a^*}{1-a^*} , \nonumber\\
	\frac{\bm}{\b} &= \frac{p^2-b^*}{1-b^*}
\label{EQ:DasEMT}
\end{align}
where the variables $a^*$ and $b^*$ correspond to the integrals
we defined as
\begin{align}
	a^* &= H_1 \nonumber\\
	b^* &= (\bm/\sm) H_2
\end{align}
as shown in Ref.~\cite{Mao2011}.

Using the asymptotic forms we obtained in
Eqs.~(\ref{EQ:H23final},\ref{EQ:Hone}), we have
\begin{align}
	\frac{\sm}{\s} &= \frac{p-1+(\sqrt{A}/2)(\bm/\sm)^{1/2}}{(\sqrt{A}/2)(\bm/\sm)^{1/2}} , \label{EQ:sDas}\\
	\frac{\bm}{\b} &= \frac{p^2-(\sqrt{A}/2)(\bm/\sm)^{1/2}}{1-(\sqrt{A}/2)(\bm/\sm)^{1/2}} .
\end{align}

Close to $p=1$, if we take $\b/\s \ll 1$, then
$(\sqrt{A}/2)(\bm/\sm)^{1/2} \ll 1$ and we have
\begin{align}
	\bm/\b \simeq 1 .
\end{align}
Plugging this back into Eq.~\eqref{EQ:sDas} we get
\begin{align}
	\frac{\sm}{\s} &= \frac{p-1+(\sqrt{A}/2)(\b/\sm)^{1/2}}{(\sqrt{A}/2)(\b/\sm)^{1/2}}
\end{align}
which is a quadratic equation in $\sm^{1/2}$, and the solution
to this leads to the same asymptotic solution as in
Eqs.~(\ref{EQ:mumAsym},\ref{EQ:muscale}).

Next we discuss asymptotic behaviors near the rigidity
threshold $\Pbtwo$ (the subscript $2$ is used to distinguish it
from the threshold $\Pb$ in EMT I). The value of $\Pbtwo$ in
EMT II can be readily obtained from
\begin{align}
	0&=p-a^* , \nonumber\\
	0&= p^2 - b^* ,
\end{align}
along with the condition that $a^* + b^*=2d/z$, as discussed in
the App.~D of Ref.~\cite{Mao2011}.  We then have
\begin{align}
	\Pbtwo = \frac{1}{2} \left(-1+\sqrt{1+\frac{8d}{z}}\right) \simeq 0.6180 .
\end{align}

The EMT II self-consistency equations~\ref{EQ:DasEMT} can then
be expanded at small
\begin{align}
	\Delta p = p-\Pbtwo ,
\end{align}
and this leads to
\begin{align}
	\sm = \s \Phi_2 \left(\b/(\s a^2)\right) (p-\Pbtwo)
\end{align}
with
\begin{align}
	\Phi_2 (x) = \frac{(2\Pbtwo+1)x}{\Pbtwo b_{b,2} + (1-\Pbtwo) x}
\end{align}
This is in the same scaling form as in EMT I
[Eqs.~\eqref{EQ:scaling-2} and \eqref{EQ:scalefn-2}], apart
from different constant factors.

%

\end{document}